\documentclass[manuscript]{aastex}
\usepackage{subfig}

\newcommand{\cF}{\mathcal{F}}

\shorttitle{Block Sequential Deconvolution}
\shortauthors{Homrighausen et al.}

\begin{document}

%\title{Block Sequential Deconvolution with Temporally Varying Kernels}
\title{Image Coaddition with Temporally Varying Kernels}
\author{Homrighausen, D; %\altaffilmark{1} 
Genovese, C%\altaffilmark{1}
}
\affil{Carnegie Mellon University, Department of Statistics.
    Pittsburgh, PA 15232}
\email{dhomrigh@stat.cmu.edu}
\and
\author{
Connolly, A; %\altaffilmark{2}
Becker, A.C; %\altaffilmark{2}
Owen, R%\altaffilmark{2}
}
\affil{University of Washington, Department of Astronomy. Seattle, WA}

\begin{abstract}
  Large, multi-frequency imaging surveys, such as the 
  Large Synaptic Survey Telescope (LSST),
  need to do near-real time analysis of very large datasets.  This raises a
  host of statistical and computational problems where standard methods
  do not work.  In this paper, we
  study a proposed method for combining stacks of images into a single
  summary image, sometimes referred to as a template. This task is commonly
  referred to as image coaddition.  
  In part, we focus on a method
  proposed in \citet{kais2004}, which outlines a procedure for combining
  stacks of images in an online fashion in the Fourier domain. 
  We evaluate this method by
  comparing it to two straightforward methods through the use of various
  criteria and simulations.  Note that the goal is not to propose these
  comparison methods for use in their own right, but to ensure that additional
  complexity also provides substantially improved performance.
\end{abstract}


\section{Introduction}
In astronomy today, the standard technique for acquiring deep
astronomical images is through a series of short, often dithered,
exposures. Applying this procedure enables the removal of cosmic rays,
the filtering of bad pixels, and the masking of device defects. It
also increases the dynamic range of the resulting image (as bright
sources in single, long exposures saturate) and provides better
control of the underlying point spread function (PSF) as long as the 
PSF is fully sampled.

To achieve the depth of a long exposure from a series of exposures
requires that we combine the individual images, accounting for the
variation in seeing, sky transparency, and background
variability. This process is typically referred to as ``image
coaddition'' and is at the heart of many image processing systems. The
criteria that we optimize in the image coaddition depends on
the science in question at hand. For example, in faint galaxy surveys
(e.g.\ the Hubble Ultra Deep Field, \citet{beck2006})
photometric accuracy might be paramount, but for weak lensing
surveys (e.g. the Canada France Hawaii Telescope Legacy Survey, 
\citet{park2007}), the preservation of the underlying image
resolution may be the primary concern. Given these differing objectives
two open questions remain; how do we optimize the coaddition of astronomical images 
and is there is an optimal criterion that is relevant to all
scientific goals?

With a new generation of deep imaging surveys coming on-line
throughout this decade, including the Panoramic Survey Telescope \&
Rapid Response System
(Pan-STARRS\footnote{http://pan-starrs.ifa.hawaii.edu}), the Dark
Energy Survey (DES\footnote{http://www.darkenergysurvey.org}), and the
Large Synoptic Sky Survey (LSST\footnote{ http://www.lsst.org}), each
producing petabytes of data, image coaddition cannot just be assed by
the statistical analysis; it is also a question of the
computational efficiency of the different approaches. In this paper,
we compare several different approaches for image coaddition,
taking into account their statistical properties and their
computational cost. We consider in detail the Fourier-domain methods
proposed in \cite{kais2004} along with several straightforward methods, such as
running means or medians. We compare these methods using 
two performance criteria: flux conservation (i.e., pixelwise 
photometric accuracy) and image resolution (using a metric we call Image Quality).
Using these comparisons we can make statements about whether the additional complexity
outweighs the extra costs they incur.  Our goal is to provide a framework
for assessing the magnitude of the practical advantages more complex
methods might provide.

In Section 2, we
set up the statistical model underlying our analysis and outline the
properties of the various techniques. In Section 3, we describe the
simulations and define the performance criteria, and, in Section 4, we
present the results of our simulations.
% Computational cost

\subsection{Mathematical Setup} \label{sec:mathematicalSetup}
Consider a single image of a particular patch of sky.
The object of interest is the true scene, which we represent
as a function $g$ that takes a two-dimensional argument corresponding to
position in the image and returns an intensity. For ground-based viewing,
however, we observe only a blurred, discretized, and noisy version of
$g$. We represent the blurring (caused by atmospheric seeing and
instrument artifacts) by an operator $K$.  In this case, $K$ is
 associated with a PSF $k$ such that 
the action of $K$ on $g$ is to integrate $g$ against $k$.
 
 Now, suppose instead of getting one image 
we get a stack of $L$ images, all of the same patch of sky
with science $g$.  We assume for this discussion that the images have
been registered. Then, due to atmospheric conditions changing over time, 
for each image $i$ there is a possibly different
operator $K_i$ with associated PSF $k_i$.  Then, if we define a regular grid of
pixel centers $x_1, x_2, \ldots$,
the observation on the $i^{th}$ image at the $j^{th}$ pixel is\footnote{To simplify
notation, we are eschewing ordered pairs for representing two dimensional
coordinates in favor of single variables. Thus we are using a linear list
of pixels and are using single two-dimensional arguments (e.g., $x$) for
points in the sky/image. Note that integrals over such two dimensional
arguments are actually double integrals.} 
\begin{equation}
Y_{ij} = \int k_i(x_j,y)g(y)\,{\rm d}y \,+\, \epsilon_{ij},
\label{eq:Fredholm}
\end{equation}
where the $\epsilon_{ij}$'s are noise random variables with suitable
distributions.  Note that equation (\ref{eq:Fredholm}) also defines
the operator $K_i$ as integrating $g$ against the PSF $k_i$.
The statistical problem is to estimate $g$ given
observation of $Y = (Y_1, Y_2, \ldots)$.

What makes the problem challenging is that the operators $K_i$ are \emph{ill
conditioned}, meaning that large changes in $g$ can make relatively small
changes in $Y_{ij}$, making recovery of $g$ from $Y_{ij}$ unstable. This is
so
because blurring dampens high frequency structure, so the singular values
of this integral operator decay quickly to zero. Thus, even small levels
of noise can critically obscure structure in $g$.

\begin{figure}[!h]
\centering
\subfloat[No Noise: Example]
{\label{fig:NND}\includegraphics[width=2.9in]{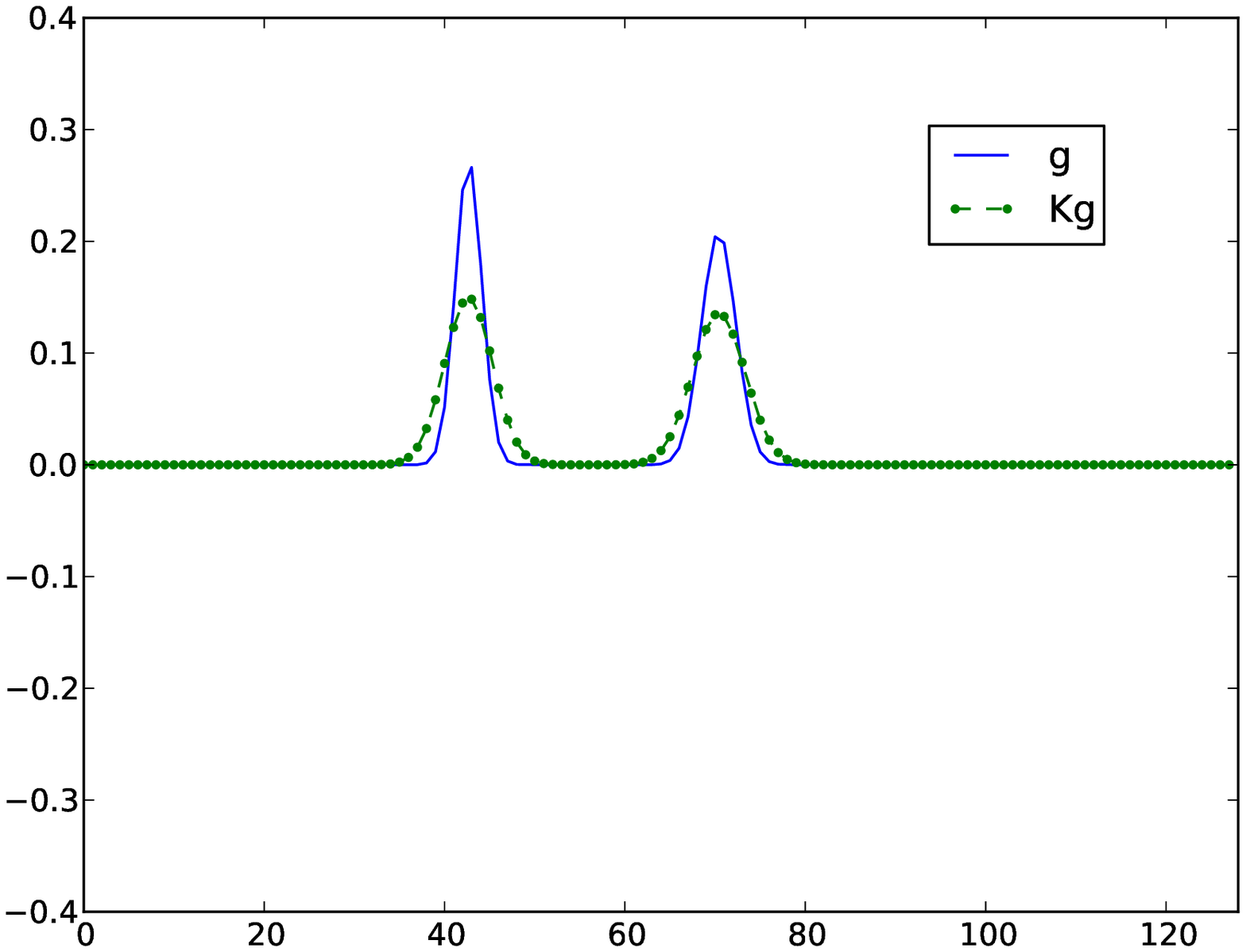}}
\subfloat[10e9:1 SNR: Example]
{\label{fig:YND}\includegraphics[width=2.9in]{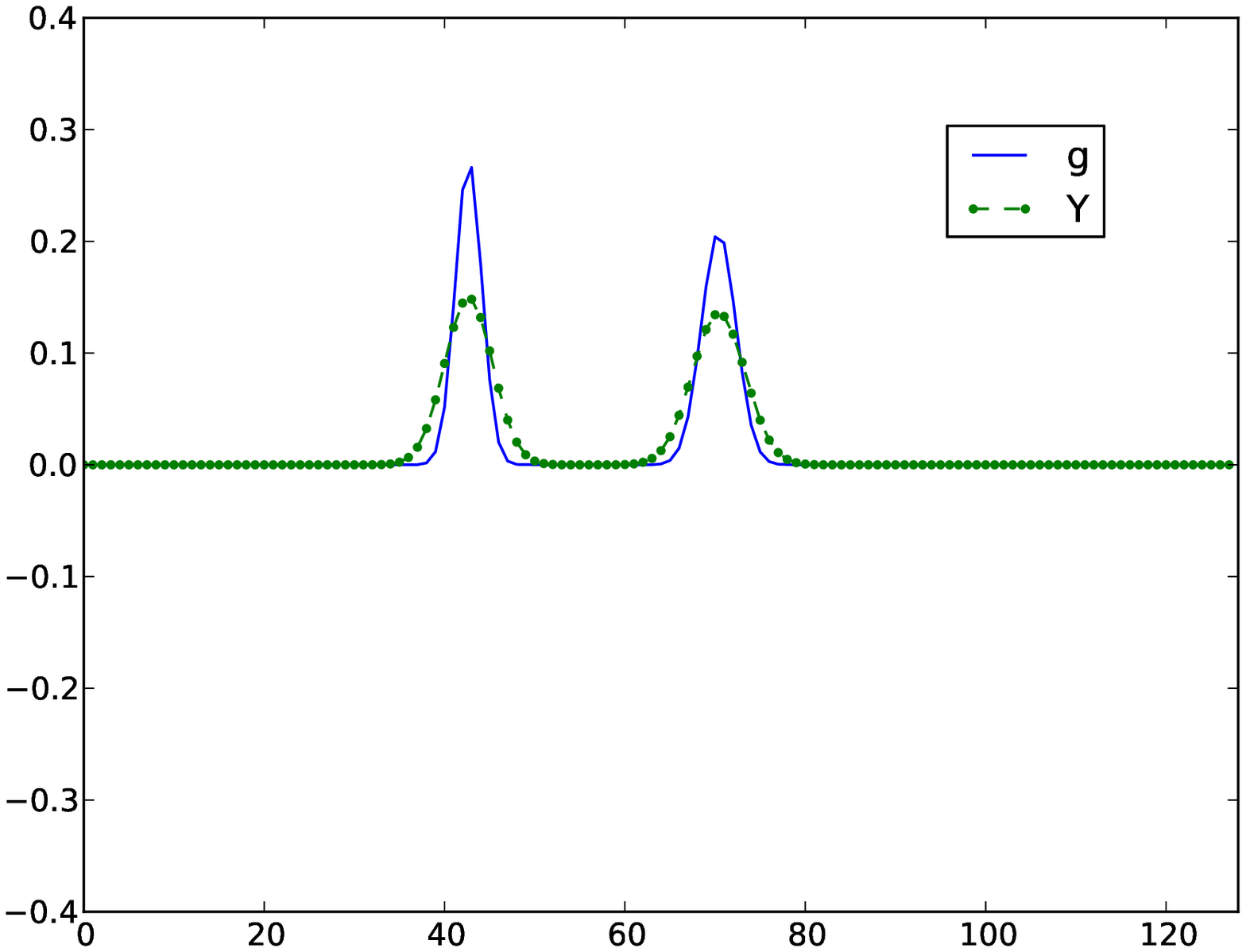}}\\
\subfloat[No Noise: Estimate]
{\label{fig:NNE}\includegraphics[width=2.9in]{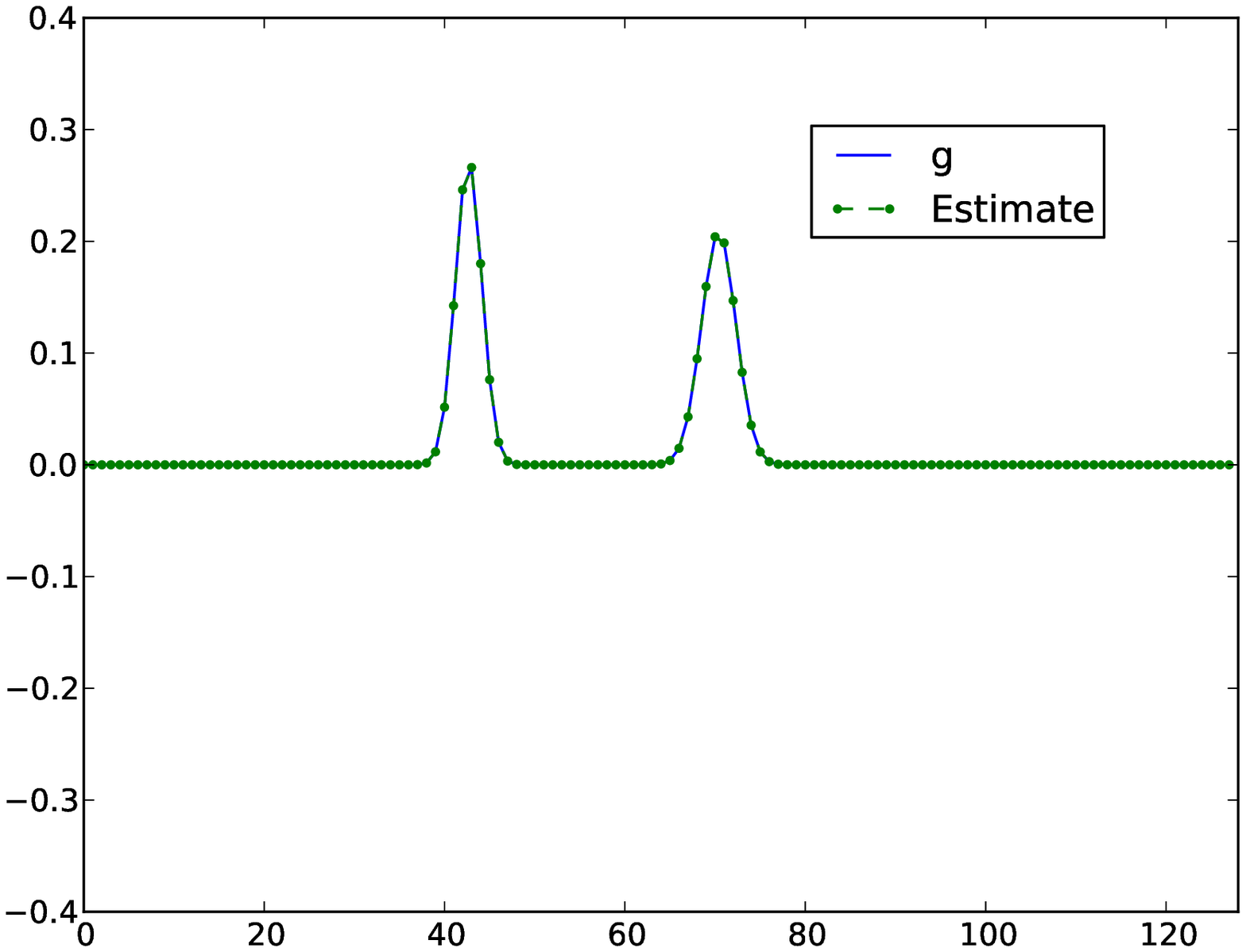}}
\subfloat[10e9:1 SNR: Estimate]
{\label{fig:YNE}\includegraphics[width=2.9in]{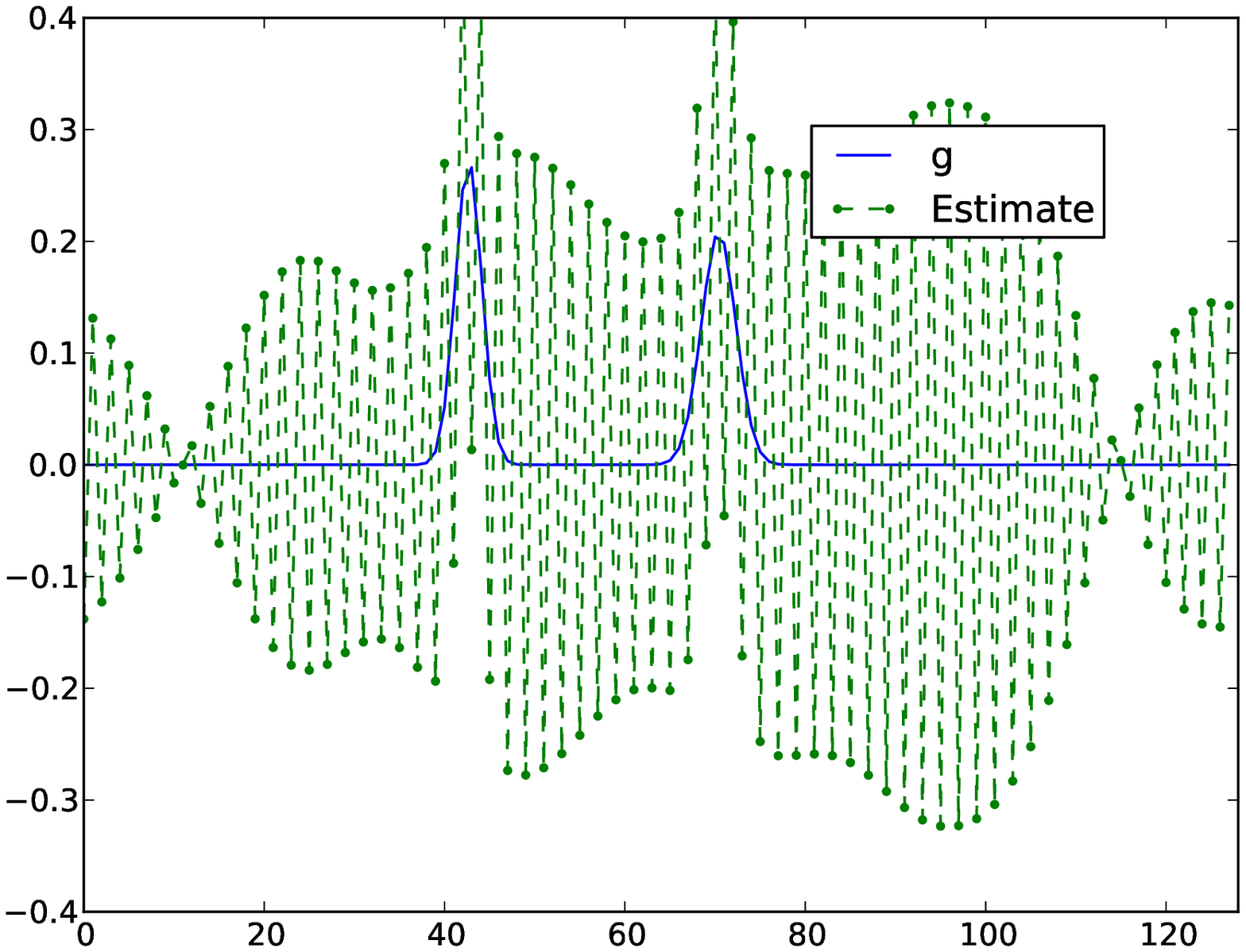}}
\caption{  We simulate a very simple case of a $g$ that is a mixture of
two Gaussians, represented as blue, solid curves.
We apply constant Gaussian seeing as well, which 
results in a new function $Kg$, displayed as a green, dashed curve. Then, we
make observations at discrete, regular points, plotted as solid green
dots. In (a) we observe $Kg$ directly and in (b) we observe $Kg$ under
a miniscule amount of noise, 10e9:1 SNR. Using
equation (\ref{eq:FourierInvOperator}),
we attempt to recover $g$ using these observations. These recoveries 
are in (c) and (d), 
respectively.  Notice that we can perfectly reconstruct $g$ in absence
of noise. However, even though $Y$ and $Kg$ are visually identical, the same
inverse approach fails spectacularly in the second case. }
\label{fig:FourierInvOperator}
\end{figure}

To see this in a simple example, assume that the PSF does not vary in
shape across the image, which makes $k(x,y) = k(x -
y)$ a convolution kernel.  It follows that by
using the Fourier transform $\cF$, we can recover $g$ exactly in the
absence of noise by 
\begin{equation}\label{eq:FourierInvOperator}
g = {\cF}^{-1}\left(\frac{\cF(Kg)}{\cF k}\right) 
\end{equation}
because $\cF(K g) = \cF k \,\cF g$. While the inverse exists in the
absence
of noise, the decay of $\cF k$ at high frequencies can produce
catastrophically high variances in the presence of noise, which produces
nontrivial and random contribution at high frequencies (so the numerator
is non-zero but the denominator goes to zero). Figure \ref{fig:FourierInvOperator} demonstrates this
effect in a 1-dimensional analog of the problem, comparing noise free
reconstruction via equation (\ref{eq:FourierInvOperator})
with reconstruction under noise with a 10e9:1 signal-to-noise ratio (SNR).
The
estimate in the latter case is wholly untenable.

\subsection{Statistical Model}\label{sec:model}
We assume that each observation is drawn from a distribution with mean 
\begin{equation}
\theta_{ij} = (K_i g)(x_j) = \int k_i(x_j, y) g(y)\,{\rm d}y.
\label{eq:theta}
\end{equation}
Because the observation process in the telescope involves counting
photons, the distribution of $Y_{ij}$ is well modeled by
a Poisson distribution \citep{Hu2007,Scully1969}. Thus, we assume that the $Y_{ij}$'s are
independent Poisson$\langle\theta_{ij}\rangle$
random variables. If the mean counts $\theta_{ij}$ are large enough,
the Gaussian approximation to the Poisson distribution is accurate, and
we can model the $Y_{ij}$'s as independent
Normal$\langle\theta_{ij},\theta_{ij}\rangle$ random variables.

The statistical problem is to form an estimator $\hat g$ of $g$ given 
observations $Y_{ij}$.  In the present context, $\hat g$ is the template
or coadd we are attempting to create.  For each $\hat g$
we need ways to measure its misfit 
relative to the true scene $g$.  In this paper we consider two. 
The first criteria corresponds to flux conservation, which we
measure through Mean Integrated Squared Error (MISE), and the second
criteria we refer to as Image Quality.  We give a brief description of
these here.  See section \ref{sec:methods} for a more thorough discussion.

Flux conservation gives an indication about
how well a method maintains the spatial location of a given amount flux.
Mean Integrated Squared Error (MISE), which corresponds
to the expected value of the integrated squared difference between any
recovered image and the true image $g$ is the measure of flux conservation
we use.
Specifically, suppose we form a template $\hat{g}$. Then
\[
MISE(\hat{g},g) = \mathbb{E} \int (\hat{g}(x) - g(x))^2 dx.
\]
See \citet{silv1985} for a thorough overview of MISE. 

\begin{figure}[!h]
  \centering
  \subfloat[Good MISE, Good Image Quality]
{\label{fig:GG}\includegraphics[width=3in]{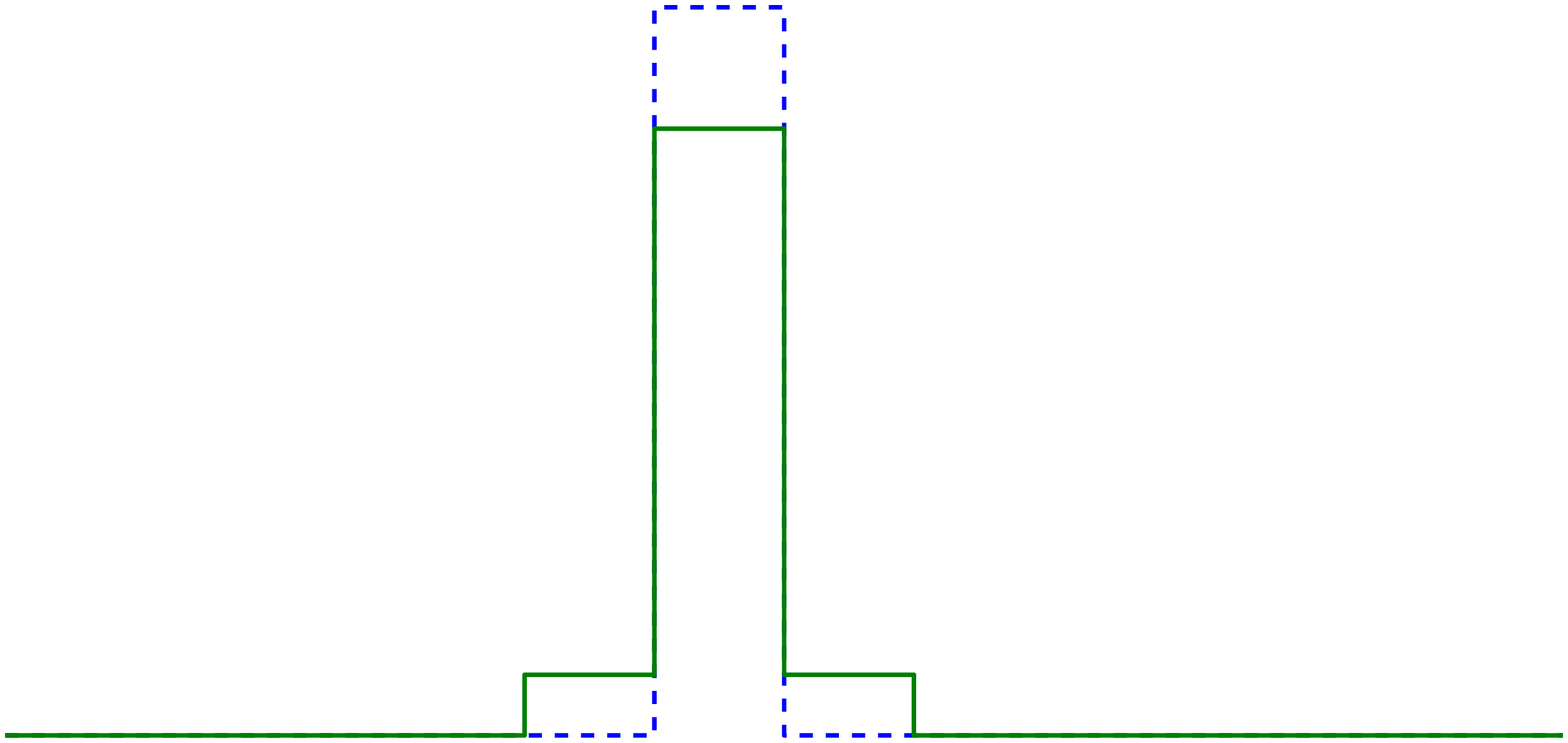}}
  \subfloat[Bad MISE, Good Image Quality]
{\label{fig:BG}\includegraphics[width=3in]{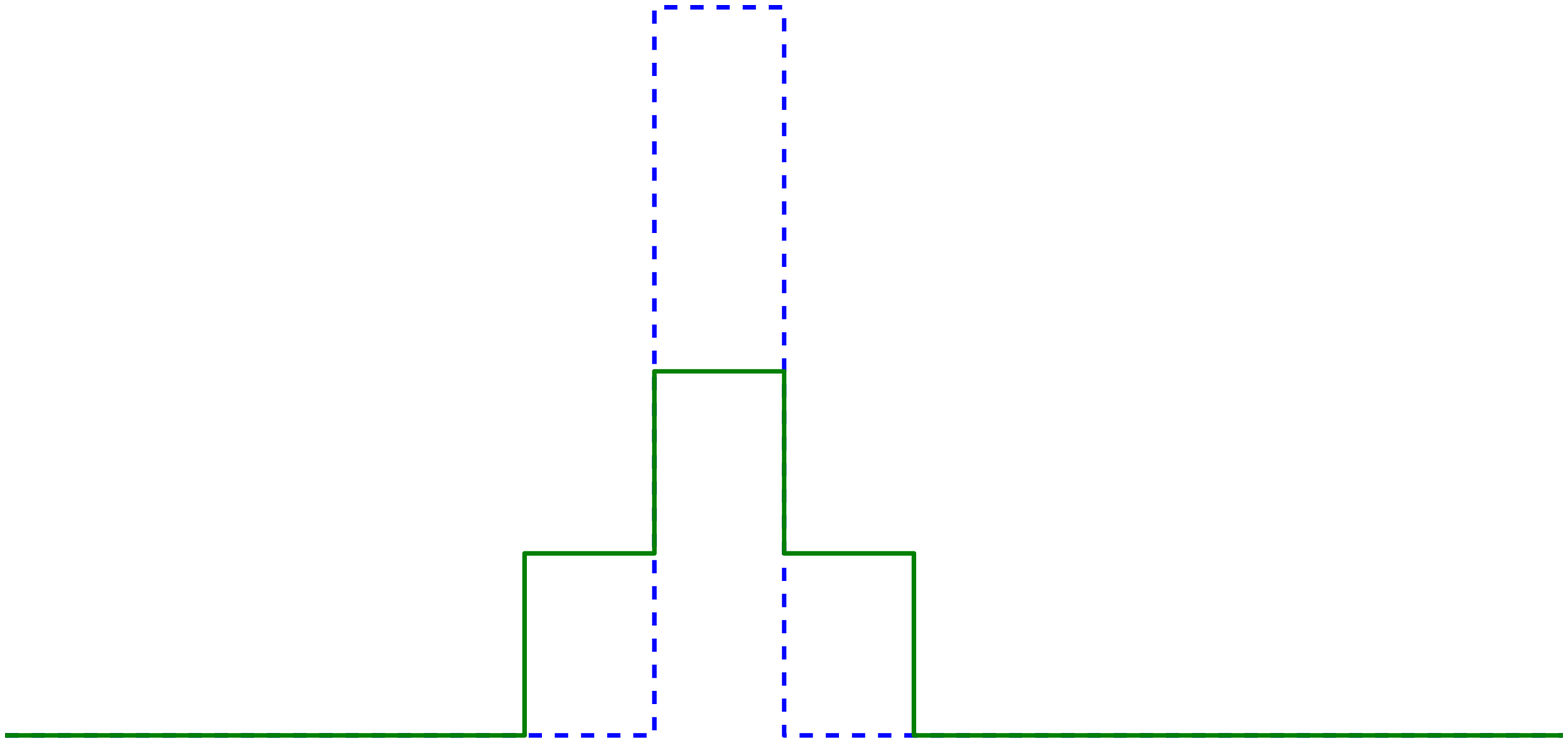}}\\
  \subfloat[Good MISE, Bad Image Quality]
{\label{fig:GB}\includegraphics[width=3in]{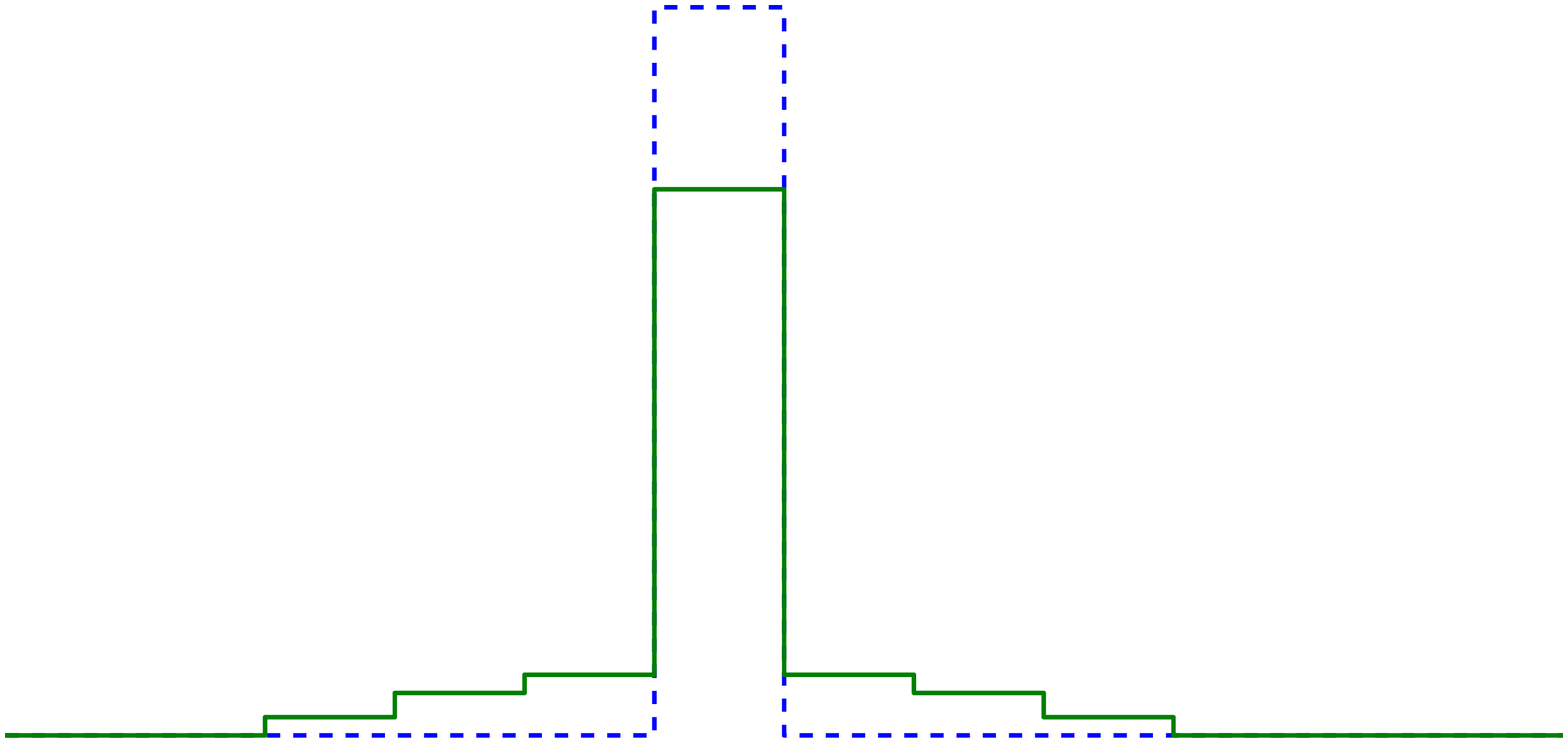}}
  \subfloat[Bad MISE, Bad Image Quality]
{\label{fig:BB}\includegraphics[width=3in]{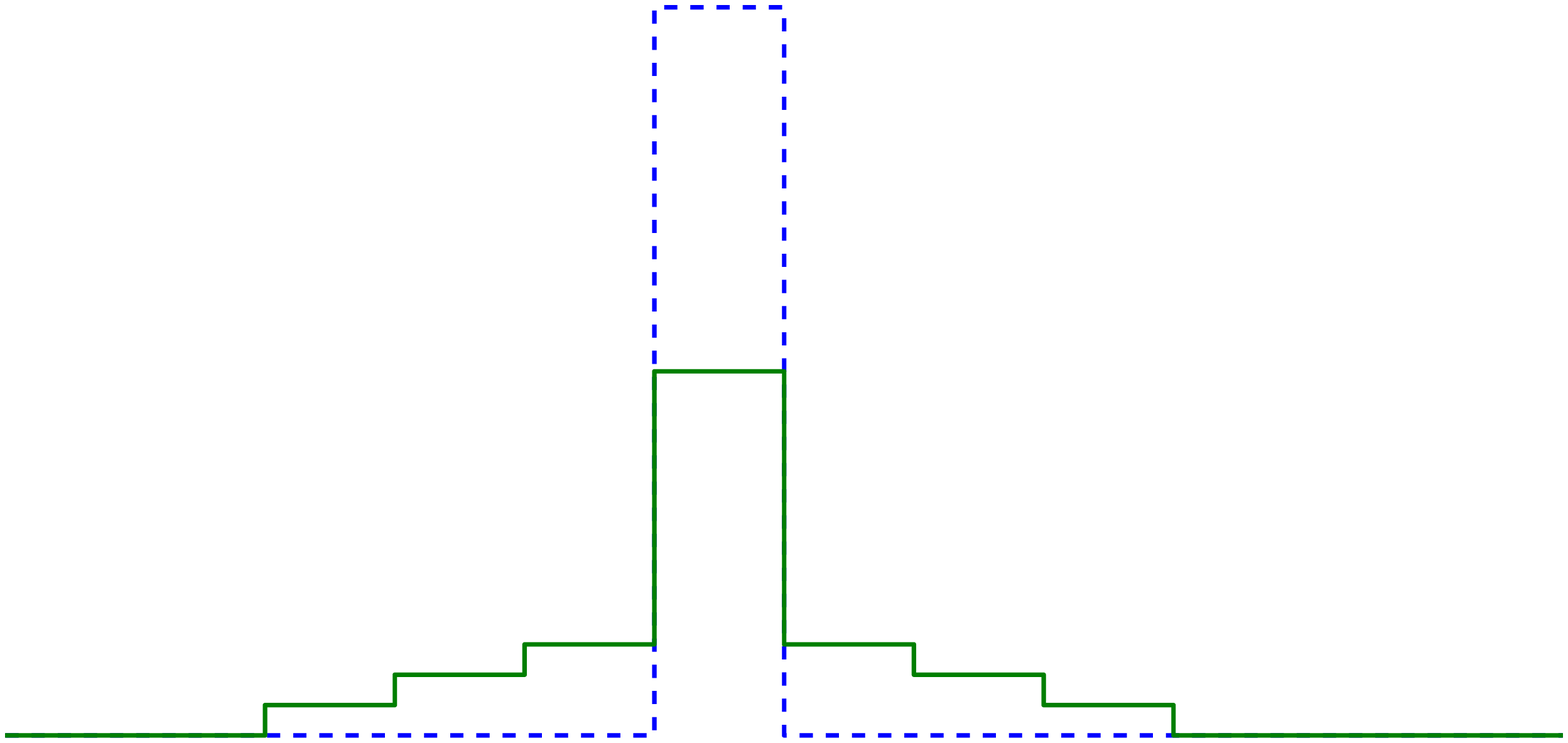}}
           \caption{The dashed line represents the single pixel
             $\delta$-function of $g$.  The solid line corresponds to
             a possible reconstruction.  The first column has
             estimates that have good MISE and the first row has
             estimates that have good Image Quality.  The second column
             and row have poor MISE and Image Quality, respectively.}
  \label{fig:miseSharpnessComparison}
\end{figure}

While MISE measures the degree to which $\hat g$ maintains flux in the correct
location, it doesn't distinguish between different scatterings of the remaining
misspecified flux.   We define a second comparison, which we call Image Quality,
that is designed to measure
how spread out the source is, as this isn't captured by MISE. For Image Quality,
we do
a very rough PSF estimation on a source in the coadded image $\hat g$.  
The relative size of these PSFs give us an indication about how spread out
an object gets by the coaddition. See Fig. \ref{fig:miseSharpnessComparison} for
an example of proceedures with combinations of good and bad MISE and Image Quality.

One significant complication in practice is that the PSFs are unknown
and must be estimated as part of the procedure. While it is ideal to
simultaneously estimate the PSF and $g$, it is common, and typically effective, to
first estimate
the PSF and then use the estimate as a 'known' PSF in the coaddition
procedure. Because our focus is on the relative performance of techniques
for estimating $g$, we will follow standard practices and assume 
that the PSFs are
known. This produces a comparison of the techniques at their best
but does not account for how well the procedures tolerate mis-estimation
of the kernels.  Note that the straightforward comparison methods do not require
PSF estimates.  So, any method that does require knowledge of the PSF would need to
be substantially better, given it has access to much more information.

\section{Methods}
In this section, we describe a variety of methods that have been applied
to the coaddition problem, with particular focus on the Fourier domain
approach we are analyzing.  Note that we restrict our attention to non-iterative
methods.

\emph{Lucky Imaging}. In lucky imaging,
a large number of images are observed and only the 'best,' according to
some criterion such as seeing, are retained. \citet{frie1978} and \citet{tubb2003} 
describe such implementations in detail.
An advantage of this approach is that the reconstruction of the true
scene
 is based entirely on high quality data. A disadvantage is that the
method
requires storing many images to determine which are best. Moreover, the
images that are discarded can contain useful information about the scene
that is, in effect, wasted.

\emph{Pixelwise Statistics}.
If a stack of images is aligned, the values at a particular pixel in
each image represent a random sample from a common distribution. The
aligned pixel stacks can thus be used to estimate the parameters of these
distributions and in turn the true scene. Basic choices would be a mean,
median, or perhaps a trimmed mean to account for heavy-tailed noise, but
more sophisticated estimators tailored to particular distributional
assumptions can be constructed. The advantages of these
approaches are computational and conceptual simplicity. Moreover, the mean
can be computed sequentially with a fixed amount of storage, although this is not
true
of the median or trimmed mean, for which the entire image stack needs to
be maintained.  A big disadvantage is that pixelwise methods tend to create
rough, discontinuous images as no information sharing is permitted between
pixels.

\emph{Fourier Deconvolution}.
In spatially constant seeing case is that
equation (\ref{eq:Fredholm}) has a convolutional structure $k_i(x,y) \equiv k_i(x - y)$.
The advantage of Fourier based methods is 
that the operator $K_i$ decomposes nicely in the Fourier domain.  An inherent
disadvantage to the Fourier approach is that great care must be used to avoid
outcomes such as Fig. \ref{fig:FourierInvOperator}.

The approach outlined in \citet{kais2004} is to Fourier
transform each image and estimate the $u^{th}$ Fourier
coefficient of $g$ by a weighted average
of the $u^{th}$ Fourier coefficient of each image.  The weighting
is accomplished so that the images with better seeing are weighted more
heavily in the average.
This method has some associated
optimality properties.  However, in practice, these properties turn out to not be useful.
See below for a more thorough description.

Note that, as presented, this method makes three nontrivial assumptions.
First, it assumes that each $k_i(x,y) \equiv k_i(x-y)$, that is, spatially constant
seeing. As we describe in Section \ref{sec:results},
while this assumption could have significant consequences when seeing
varies spatially, in practice, it does not appear to cause much problem.
The second assumption is that the $\theta_{ij}$'s are large enough
that the Gaussian approximation to the Poisson is accurate across the image.
The third assumption is that the variance of the Gaussian is a \emph{known}
value $\sigma^2_i$ that depends only on the image $i$.
These last two assumptions make the analysis easier,
and while they are often reasonable, they need not hold in practice.

To define the Fourier Deconvolution estimator, first expand $g$ into
the Fourier basis, which we write as $(\phi_u)$,
\begin{equation}
g = \sum_{u} \tilde g(u) \phi_u.
\label{eq:seriesEstimatorGeneral}
\end{equation}
In general,  we use the notation $\tilde{f}$ for the Fourier transform
of the function $f$.

Using the above assumptions, and the resulting form of
the $\theta_{ij}$'s, \citet{kais2004} proposes two estimators 
of $\tilde g(u)$.  See 
Table 1 %\ref{tab:estimatorsMoments} 
for these estimators and
some of their cumulants.
The first estimator in the table, $\widehat{ \tilde{g}(u) }$, has the smallest
variance of all unbiased estimators of $\tilde g(u)$.  Also, 
the resulting estimator of $g$, defined to be
\begin{equation}
\hat{g} = \sum_u \widehat{\tilde{g}(u)} \phi_u,
\label{eq:unbiasedEstimator}
\end{equation}
is the best linear unbiased estimator of $g$.

\begin{table}[!h]
  \begin{center}
  \begin{tabular}{lcl|lcl}
    \multicolumn{3}{c}{ Estimator} & \multicolumn{3}{c}{Expectation \& Variance} \\
    \hline
    \hline
    \multicolumn{3}{c|}{} & \multicolumn{3}{|c}{} \\
    $\widehat{\tilde{g}(u)}$ & $=$ &
    $\frac{\left(
      \displaystyle\sum_{l=1}^L \frac{\overline{\tilde{k}_l(u)} \tilde{Y}_l(u)}{\sigma_l^2}
      \right)}
    {
      \left(
      \displaystyle\sum_{i=1}^L \frac{ |\tilde{k}_i(u)|^2}{\sigma_i^2}
      \right)
    }$
    &$\mathbb{E}_g\left[\widehat{\tilde{g}(u)} \right]$  & $=$ & $\tilde{g}(u)$ \\
    \multicolumn{3}{c|}{} & $\mathbb{V}_g\left[\widehat{\tilde{g}(u)} \right]$  & $=$ &
    $\frac{\displaystyle 1}{\left(\displaystyle\sum_{i=1}^L\frac{|\tilde{k}_i(u)|^2}{\sigma_i^2}\right)}$ \\
    \multicolumn{3}{c|}{} & \multicolumn{3}{|c}{}   \\
    \hline
    \multicolumn{3}{c|}{} & \multicolumn{3}{|c}{}   \\
    $\widehat{\tilde{g}_*(u)}$ & $=$ &
    $\frac{\left(
      \displaystyle\sum_{l=1}^L \frac{\overline{\tilde{k}_l(u)} \tilde{Y}_l(u)}{\sigma_l^2}
      \right)}
    {
      \left(
      \sqrt{ \displaystyle\sum_{i=1}^L \frac{ |\tilde{k}_i(u)|^2}{\sigma_i^2}}
      \sqrt{ \displaystyle\sum_{j=1}^L \frac{1}{\sigma_js^2}}
      \right)
    }$
    &$\mathbb{E}_g\left[\widehat{\tilde{g}_*(u)}\right]$ & $= $ & $k^{*}(u)\tilde{g}(u)$ \\
    \multicolumn{3}{c|}{} & $\mathbb{V}_g\left[\widehat{\tilde{g}_*(u)} \right]$ & $=$ &
    $\displaystyle \sum_{i=1}^L \frac{1}{\sigma_i^2}$ \\
    \multicolumn{3}{c|}{}  & \multicolumn{3}{|c}{}  \\
    \hline
%    \label{tab:estimatorsMoments}
  \end{tabular}
\end{center}
\caption{
  $\tilde{k}^*(u) := \frac{  \sqrt{ \displaystyle\sum_{i=1}^L | \tilde{k}_i(u) |^2/\sigma_i^2} }
  { \sqrt{\displaystyle\sum_{j=1}^L 1/\sigma_j^2 } }$, which is the result of a
  variance stabilizing transformation.}
\end{table}

However, as alluded to previously, 
these optimality properties are not useful.
To see this, observe that the variance of $\hat g$ is the sum of the variances of $\widehat{ \tilde{g} (u)}$.
Also, since the $K's$ are smoothing operators, $|\tilde k (u)|$ is small for large $|u|$.
In fact, the worse the seeing, the smaller $|\tilde k (u)|$ becomes and can be
smaller than machine error in many cases.  Hence, as $\mathbb{V}_g[\widehat{ \tilde{g} (u)}]$
depends on the reciprocal of $|\tilde k(u)|$, the variance of $\hat g$ can be arbitrarily large.
Note that we have seen a case of this already in Figure 
\ref{fig:FourierInvOperator} where this large variance property 
causes unusable results.

\begin{figure}[!h]
  \begin{center}$
    \begin{array}{@{}c@{}c@{}c@{}}
      \includegraphics[width=2.0in]{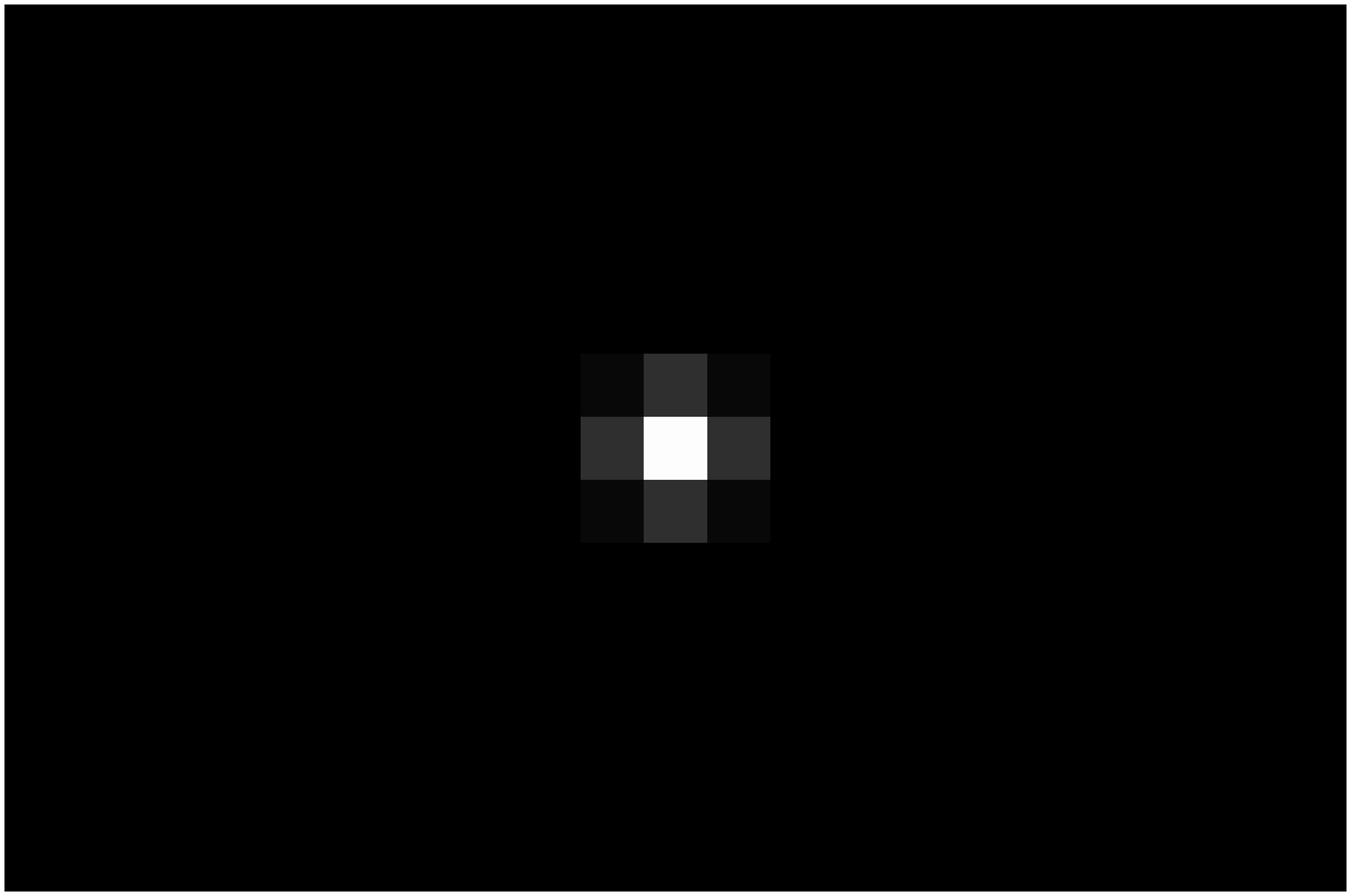} &
      \includegraphics[width=2.0in]{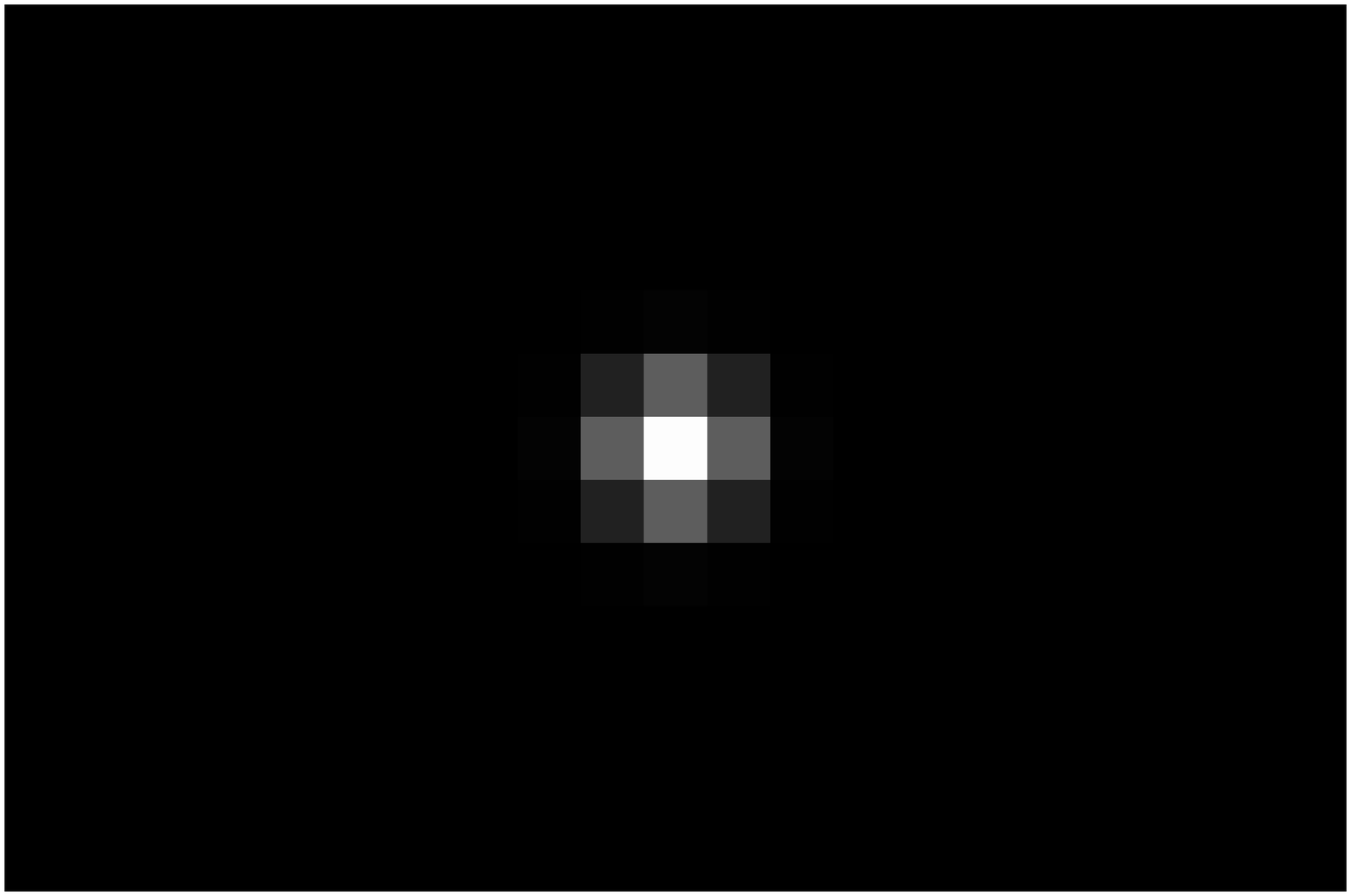} &
      \includegraphics[width=2.0in]{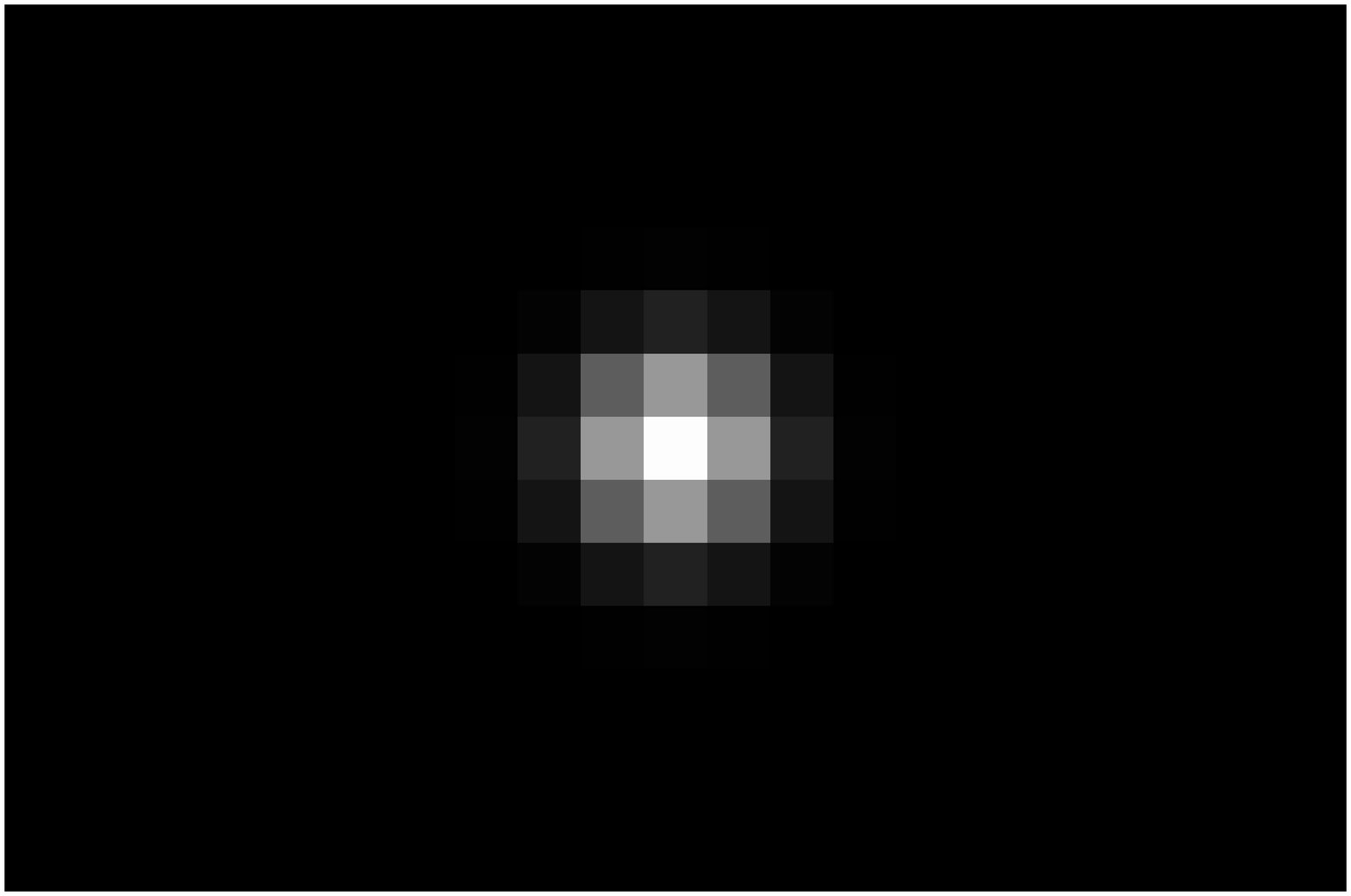} \\
      \includegraphics[width=2.0in]{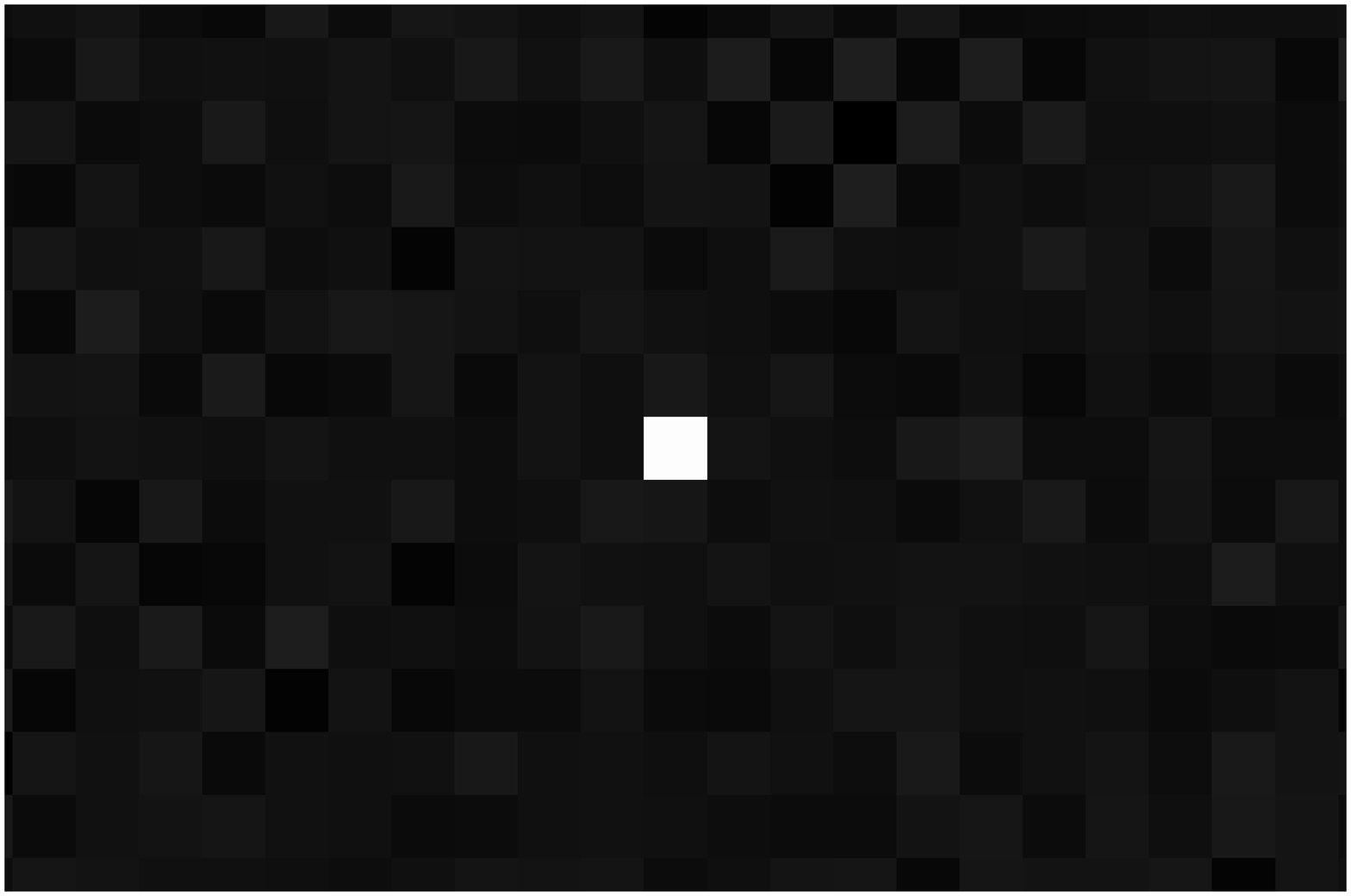} &
      \includegraphics[width=2.0in]{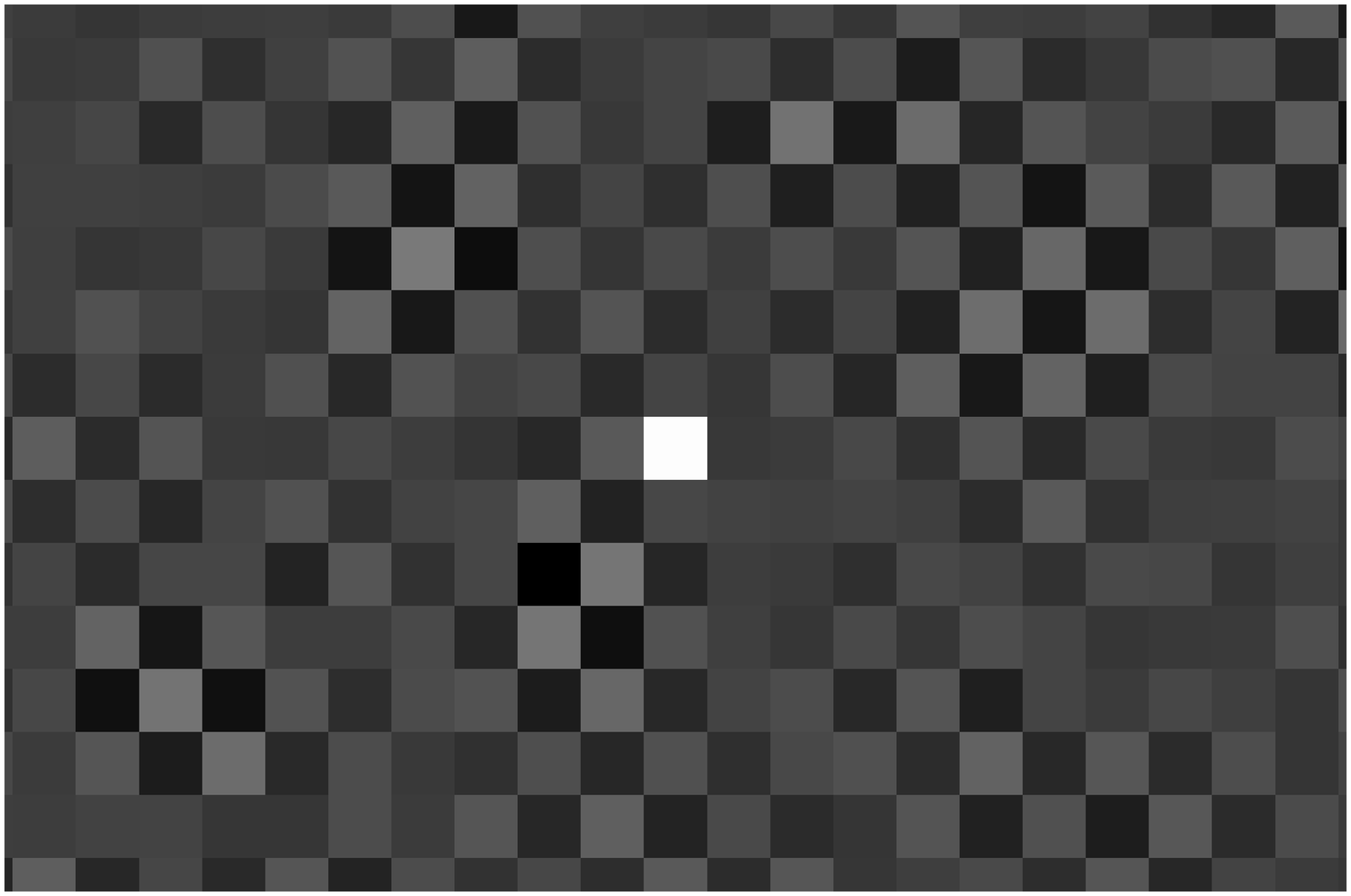} &
      \includegraphics[width=2.0in]{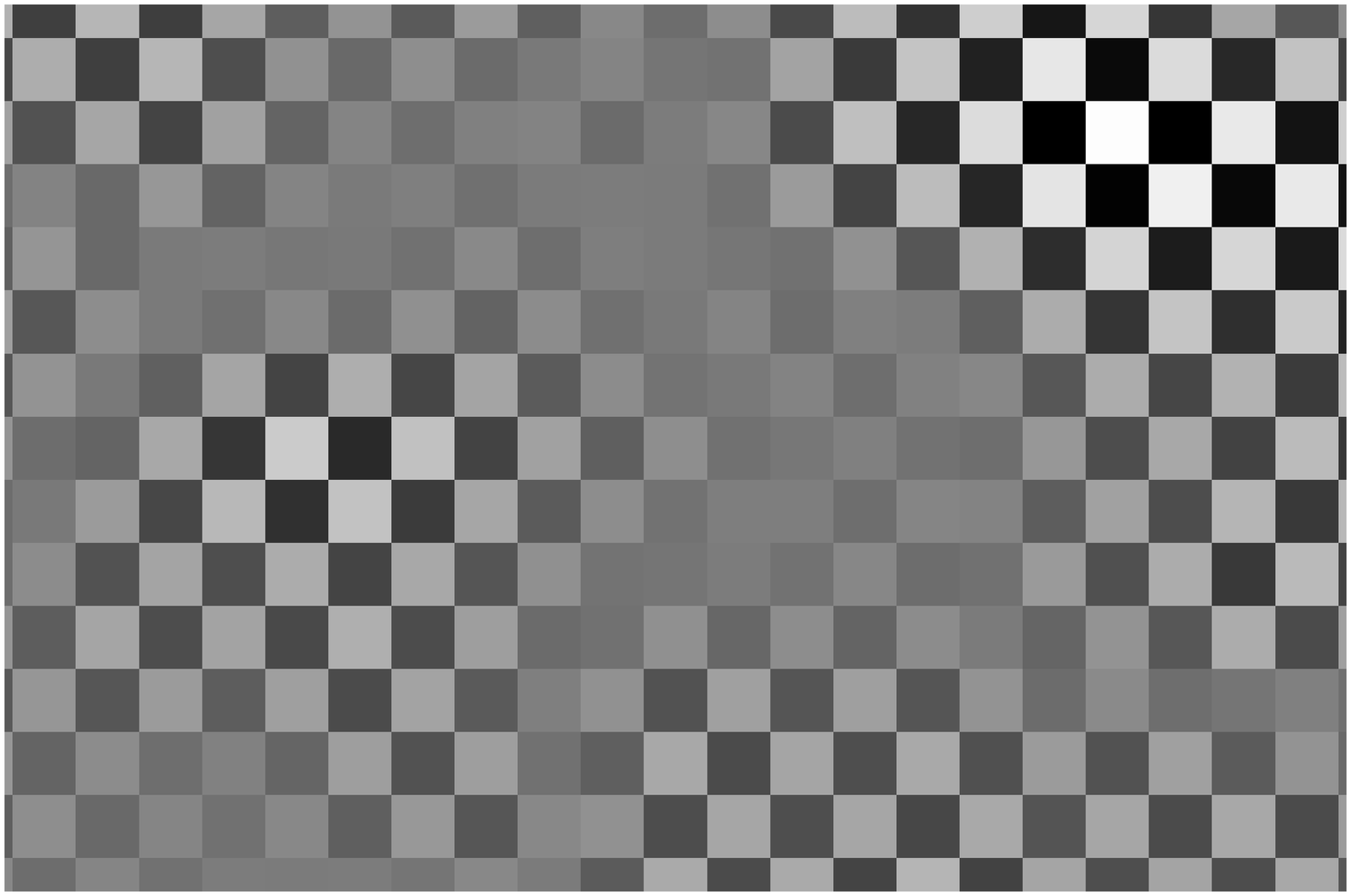} \\
    \end{array}$
  \end{center}
  \caption{This is an example of the use of $\hat{g}$.
  The true scene $g$ is a one pixel delta function representing a star (not
pictured).
  \textbf{Top Row:} An example of an input, or observed, image.
  \textbf{Bottom Row:} Reconstruction using $\hat{g}$.
  \textbf{Left to Right:} We increase the FWHM
   from 0.5 pixels to 1.1 pixels.  Note that
  this means that some of the images have an undersampled PSF.  This isn't
an issue in this
  case as we are applying the seeing ourselves and can without loss of
generality assume
  the star is at the centroid of the pixel spike in $g$.
  The performance decays rapidly as the width of the seeing increases.  In
particular,
  even the  mild amount of seeing in the right most column makes the
  estimator unusable.  The magnified high frequency information has
swamped the signal we wish to recover and left checkerboard pattern seen
in the bottom right image..}
  \label{fig:FourierEstimatorOptimal}
\end{figure}

Generally speaking, demanding  unbiased estimators is less effective
than choosing an estimator with some bias but smaller variance.
In this case, the distinction is crucial because the unbiased estimator is
unusable.
For the interested reader, Chapter 7 of \citet{wass2006} provides a good
introduction to this surprisingly broad and deep issue\footnote{Better 
performance under many different criteria comes from
giving up some bias for improved variance.  One well used example of this
compromise is Tikhonov regularization.}.
See Figure \ref{fig:FourierEstimatorOptimal} for an example of 
equation (\ref{eq:unbiasedEstimator}) in a simple simulated situation.
The ability of $\hat{g}$ to reconstruct even a simple
$g$ decays rapidly as the full width at half maximum\footnotemark, or FWHM,
\footnotetext{The full width at half maximum is defined to be
            $\textrm{FWHM}(k) := \sup_{x_1,x_2 \in P_k} |x_1 - x_2|$
            where $P_k = \left\{x : k(x) = \frac{||k||_{\infty}}{2}
\right\}$.}
 of the PSF increases.  In particular,
even the  small amount of seeing (FWHM of 1.1 pixels)
in the right most column makes the
estimator unusable\footnote{The fact that this PSF is undersampled is
not important as we numerically applied a known PSF to
a known image.}.  

\citet{kais2004} provides a correction that prevents the variance
from exploding  for large $|u|$ by
multiplying $\widehat{\tilde{g}(u)}$
by a term that is proportional to the reciprocal of its 
standard deviation. See the second row of Table 1.  This multiplication
 biases the estimator 
but bounds the variance.  We define
\begin{equation}
\hat{g}_* = \sum_u \widehat{\tilde{g}_*(u)} \phi_u
\label{eq:FourierDeconvolutionMethod}
\end{equation}
and we spend the balance of the
paper investigating its properties. We refer to $\hat{g}_*$ as the
Fourier Deconvolution (FD) method.
As an aside, since the correction 
corresponds to multiplication in Fourier space, we see that the bias
of $\hat{g}_*$ is given by $(k^**g) - g$, where the Fourier transform
of $k^*$ is
defined in the caption of Table 1 and $(a*b)$ is the convolution of $a$ and $b$.

\subsection{Computational and Space Complexities}
The computational complexity of the FD method is dominated by the need to
Fourier transform the data and kernel.  If we suppose the images are $m$ by $n$
and $N := mn$, 
then we can perform a Fourier transform in $O(N\log N)$ FLOPs\footnote{
FLOPs stands for Floating-point Operations and corresponds loosely to
additions and multiplications.} via the fast
Fourier transform.  This is the dominating computation in the FD method.  However,
it also has several order $O(N)$ computations\footnote{The square root
operation takes a variable number of FLOPs.  This is the cost for updating
the method with each new image.  The exact number depends on
the specific computing floating point unit.}.  The mean method requires $2N$
FLOPs and the median method requires $N$ comparisons to
update for each new image.  

The FD method must maintain two $N$ pixel images and the mean method 
requires one $N$ pixel image, each of which get updated after each new image
and kernel is recorded.  The median method requires the entire stack must be maintained
at all times, though not necessary in RAM.  Hence, after $L$ images have been
observed, all $L$ images must be kept for possible updating of the median.

\section{Methods: Images and Evaluation Criteria}
\label{sec:methods}
\begin{figure}[!h]
  \begin{center}
    \includegraphics[width=4.7in]{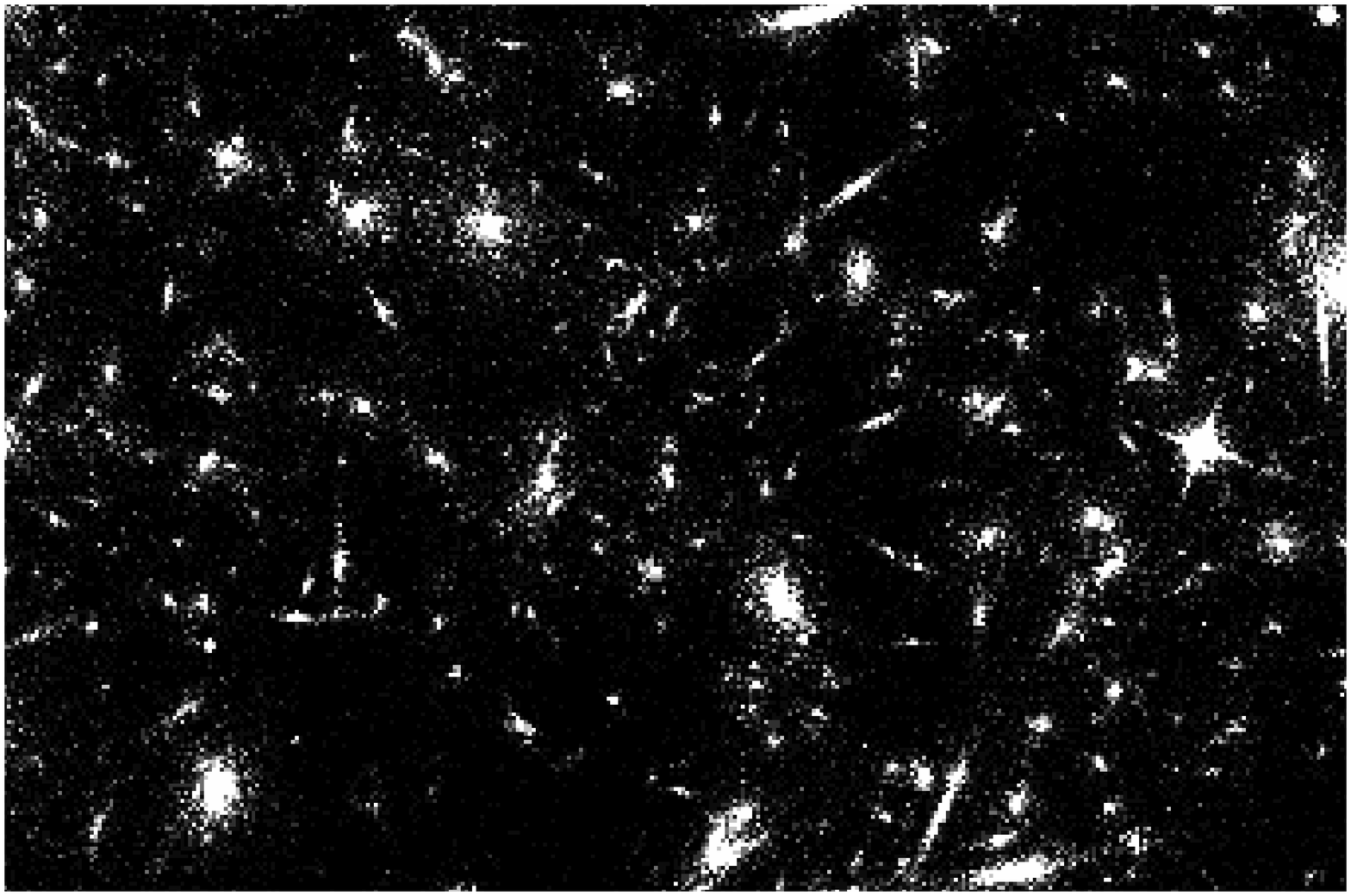}
  \end{center}
  \caption{The simulated image after being z-scaled and power transformed.   
    The pixel width of this simulation is 0.2 arcsec (0.2'').}
  \label{fig:simulation}
\end{figure}

For our evaluation metrics, we need to know the true
sky before any distortions.  Hence, we simulate an 
idealized view of the sky (above the atmosphere) using
a catalog of point and extend sources. Extended sources are
represented by Sersic profiles and the density of point
and extended sources is designed to match observations to a depth of
r$\sim$28. Figure \ref{fig:simulation} shows a representation of one
such image. From these true scenes, we apply a blurring operator $K$
and noise with variance $\sigma^2$ to represent the effect of the
atmosphere, telescope, and instrument.

Under the assumptions of the FD method,
we assume the PSFs are convolutional (by that we mean spatially constant seeing), 
but this leaves open the specific functional form for the
kernel $k$. We choose to set $k$ to be the two dimensional
Gaussian probability density function.  Our results are, however,
robust against more complicated parametrization of $k$ (e.g., a mixture of
Gaussians).

For evaluating the effectiveness of the FD method,
we compare it to the pixel-wise mean and
pixel-wise median.  Our goal is to assess the value added by
the FD method and evaluate whether the improved
performance overcomes the added cost and complexity of the method.
The mean and median procedures are  intended
as extreme comparisons rather than serious proposals and we aren't endorsing
their use in this paper.
However, if the value added of more sophisticated methods relative to even these
simplistic methods
is not great, then it suggests that simple methods may suffice in practice.
Note that the computational complexity of the methods is not a purely
academic concern. In large imaging surveys such as the LSST, 
a near constant data stream requires efficient, near real-time 
processing, and the cost of generating template images is an important factor.

For this comparison, we use the two general criteria introduced in section
\ref{sec:model}
corresponding to MISE and Image Quality. In principle, at least two 
different errors can be made when comparing an estimated template $\hat g$
to the true scene $g$.  These errors can most easily be thought of
in the context of a point source such as a star.  Our estimated image
can remove some of the flux from the point source and/or it can smooth
the image (degrading the photometric accuracy and the resolution of
the resulting template).
Figure \ref{fig:miseSharpnessComparison} shows examples of a pixelated
point source that has been reconstructed by four made-up methods.

In the MISE comparison we compute the
expected value of the integrated squared difference between the true
image and our estimate, assuming zero background.  Under the non-Gaussian
distributions, a larger variance results in more total flux in the image.
This, in turn, results in a large MISE.  To compensate for this increase,
and to make all the cases more directly comparable, we
rescale the MISE by the total signal of each image to 
compensate for this additional flux.

For the evaluation of Image Quality we fit a two dimensional, 
spherical\footnote{By spherical, we mean that the covariance matrix is
proportional to the identity matrix.}
Gaussian density to a source using least squares.  The covariance
matrix of this fitted Gaussian is then used as a metric for the width
of the point source.

\section{Results}\label{sec:results}
In this section, we present the results of our simulations.
The section is divided into two parts,
corresponding to the two evaluation criteria outlined above.

\subsection{MISE}\label{sec:MISE}

When looking at equation (\ref{eq:Fredholm}) we see two major influences on the
quality of the observation.  One is the type and severity of the
seeing, determined by the PSF $k$.  The other is the distribution of
the noise $\epsilon$.  In this section, we look at several different
scenarios for comparing the mean or median to the FD
method by changing these two factors.

\subsubsection{Increasing FWHM Comparison}
Overall, we wish to understand the impact of seeing and noise on the
performance of the methods.  To do this, we simulate a stack of images
with each image having a FWHM drawn from a distribution with mean,
$\mu_{FWHM}$.  We do this procedure for $\mu_{FWHM}$ ranging from 2
pixels to 7 pixels (0.4'' to 1.4'').  This interval is chosen so
that it contains the
expected seeing width for the LSST.  This procedure creates a sequence of
stacks of images.  

\begin{figure}[!h]
  \centering
  \subfloat[Gaussian, Low Noise]{\label{fig:lowGaussian}
\includegraphics[width=3in]{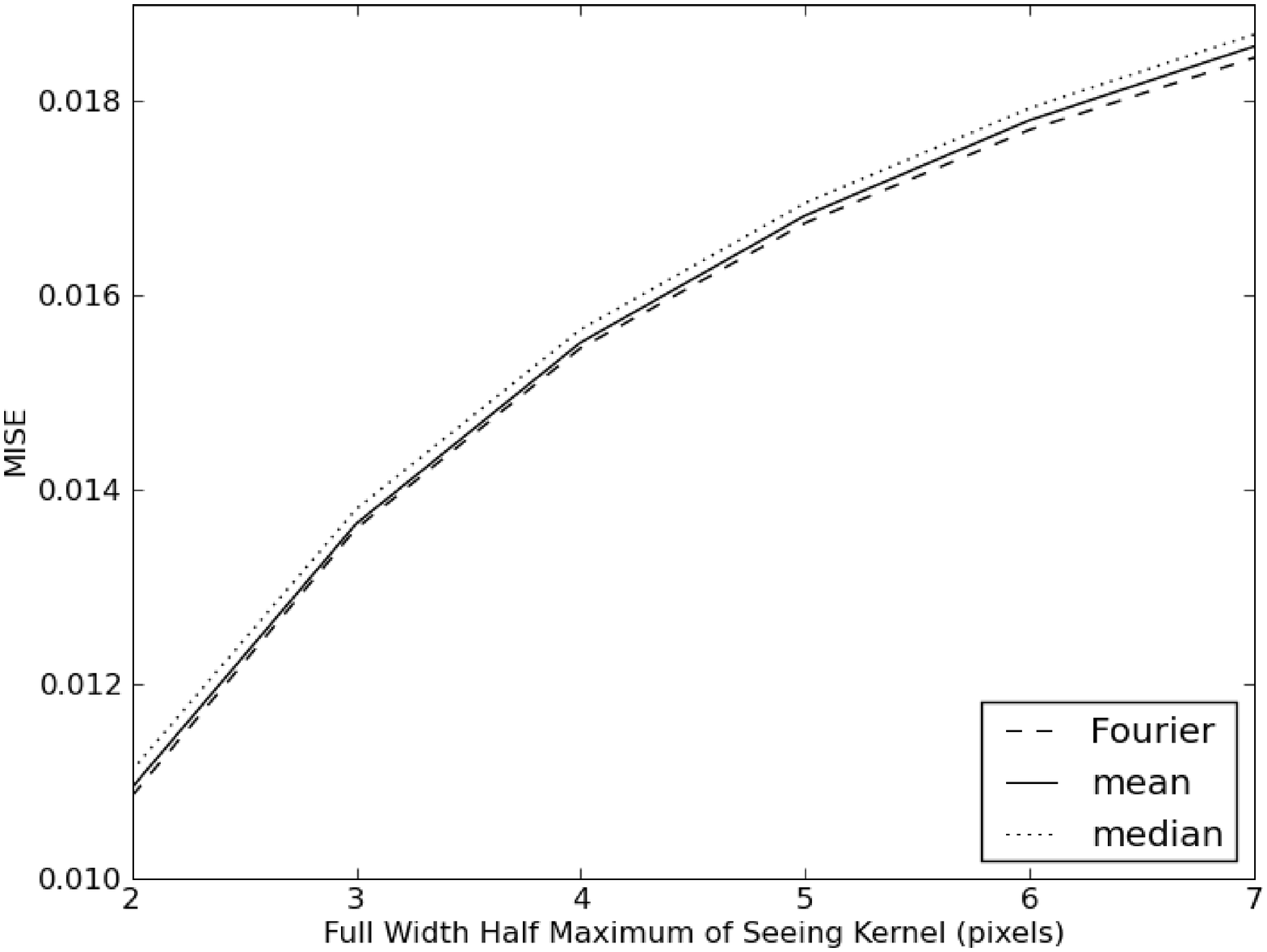}}
  \subfloat[Gaussian, High Noise]{\label{fig:highGaussian}
\includegraphics[width=3in]{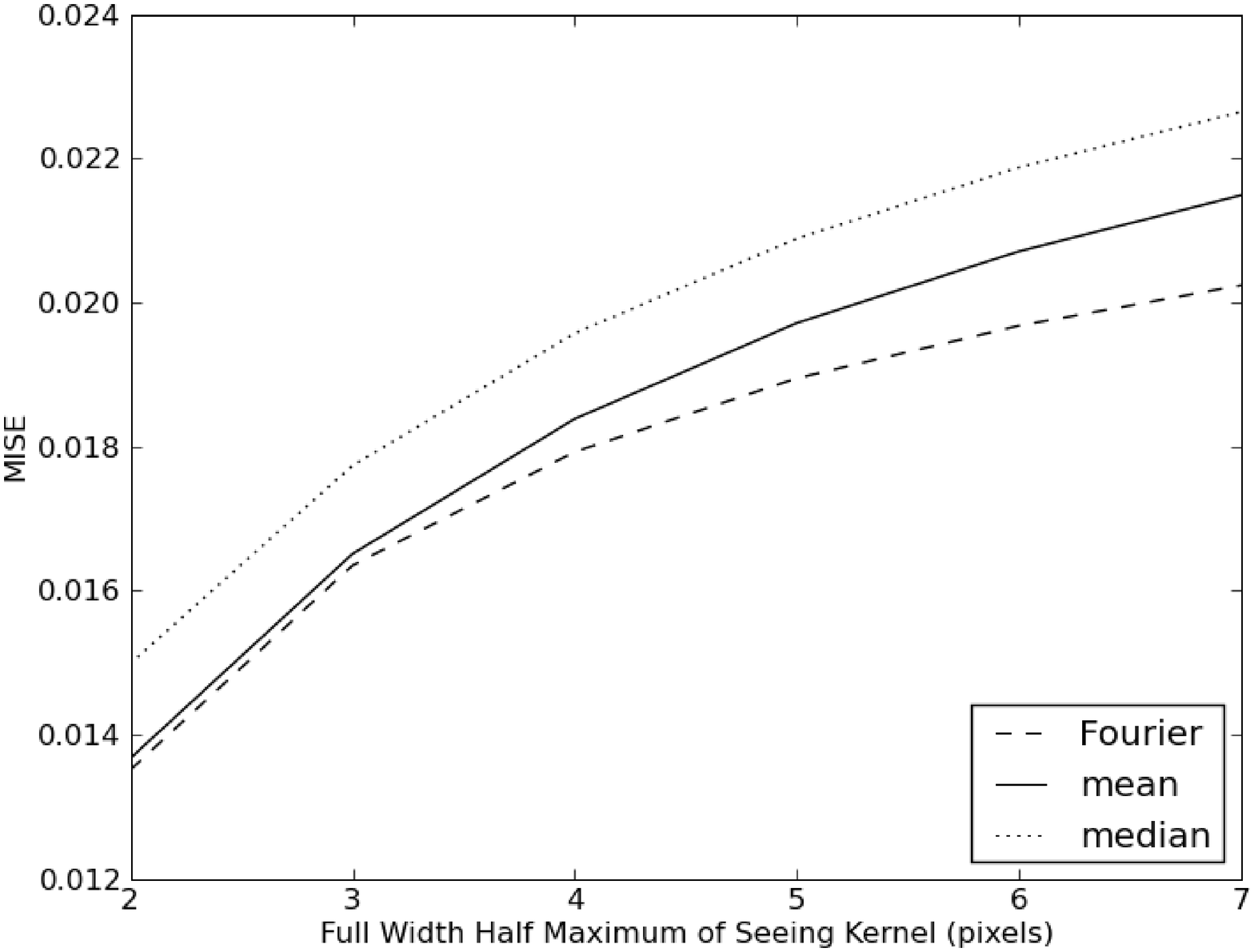}} \\
  \subfloat[Shot, Low Noise]{\label{fig:lowShot}
\includegraphics[width=3in]{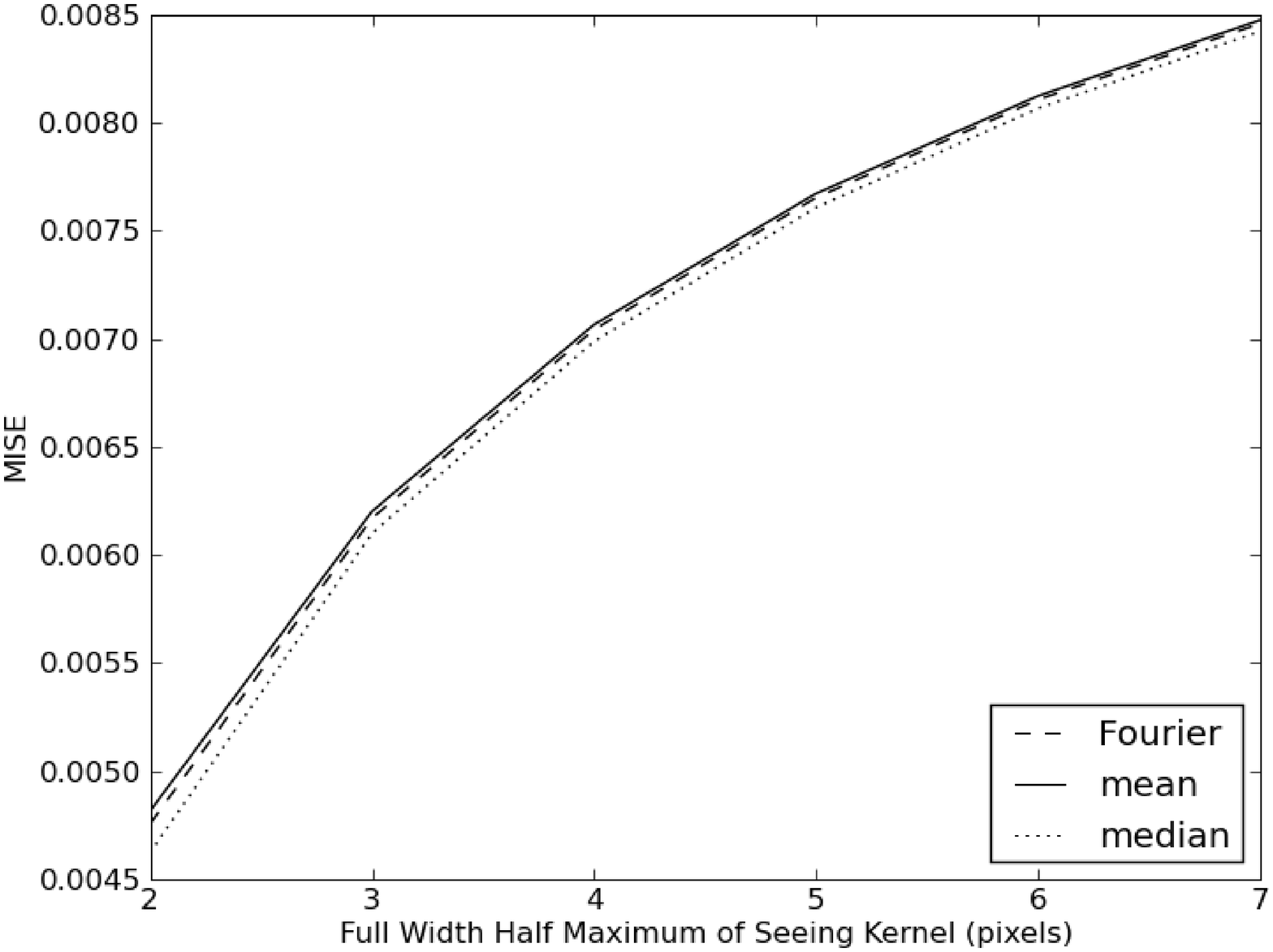}}
  \subfloat[Shot, High Noise]{\label{fig:highShot}
\includegraphics[width=3in]{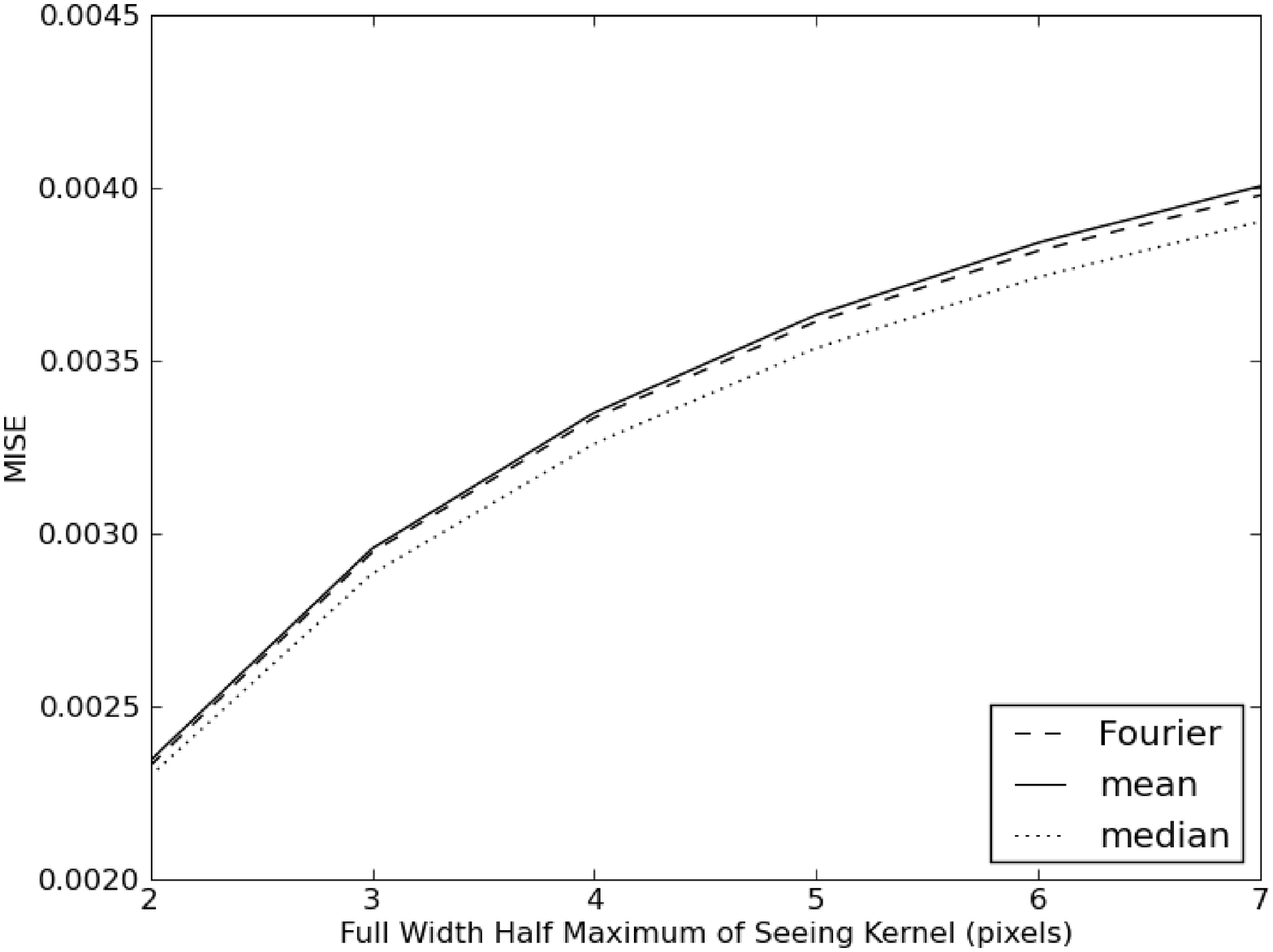}} \\
  \subfloat[Inhomogeneous Poisson, Low Noise]{\label{fig:lowPoissonIntensity}
\includegraphics[width=3in]{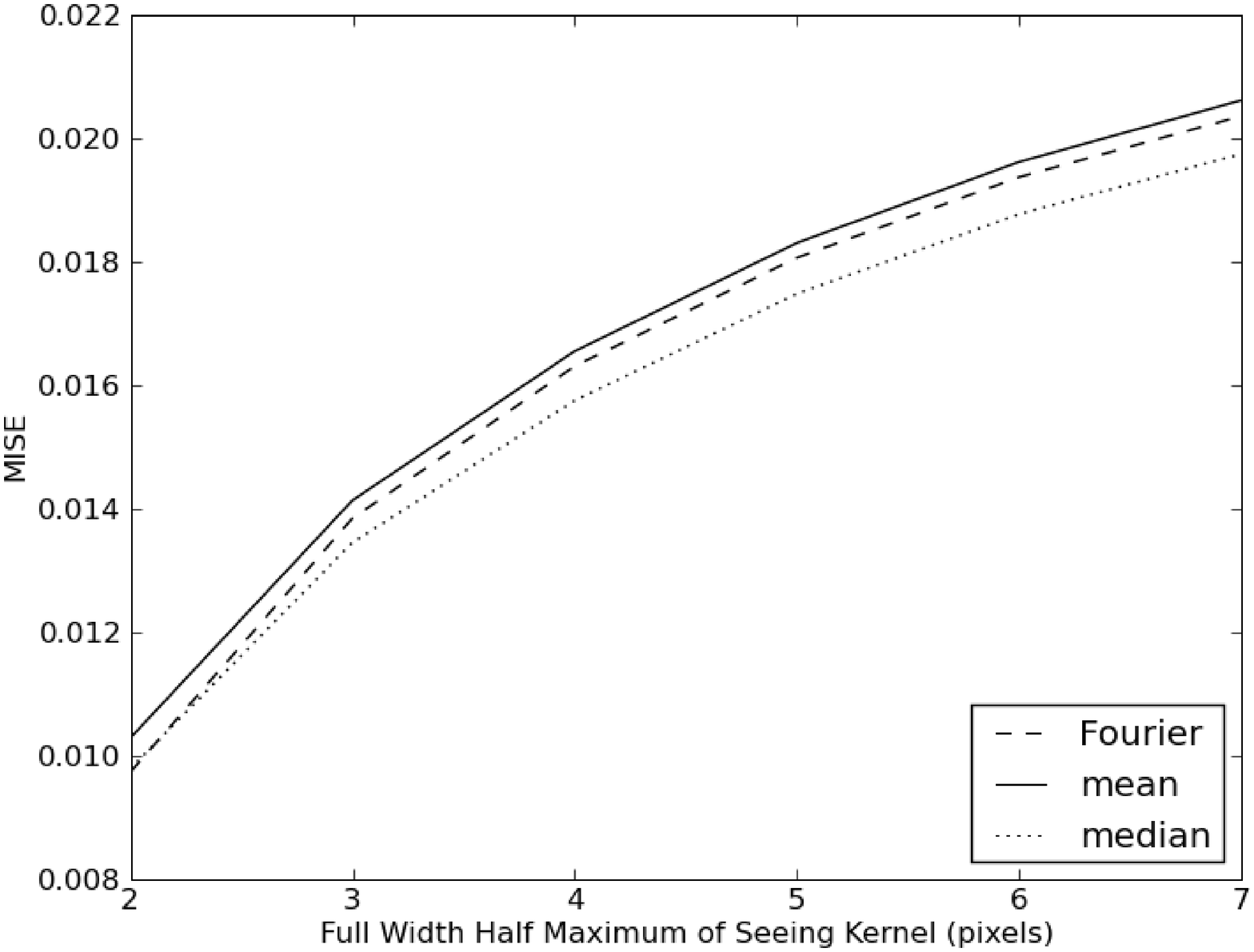}}
  \subfloat[Inhomogeneous Poisson, High Noise]{\label{fig:highPoissonIntensity}
\includegraphics[width=3in]{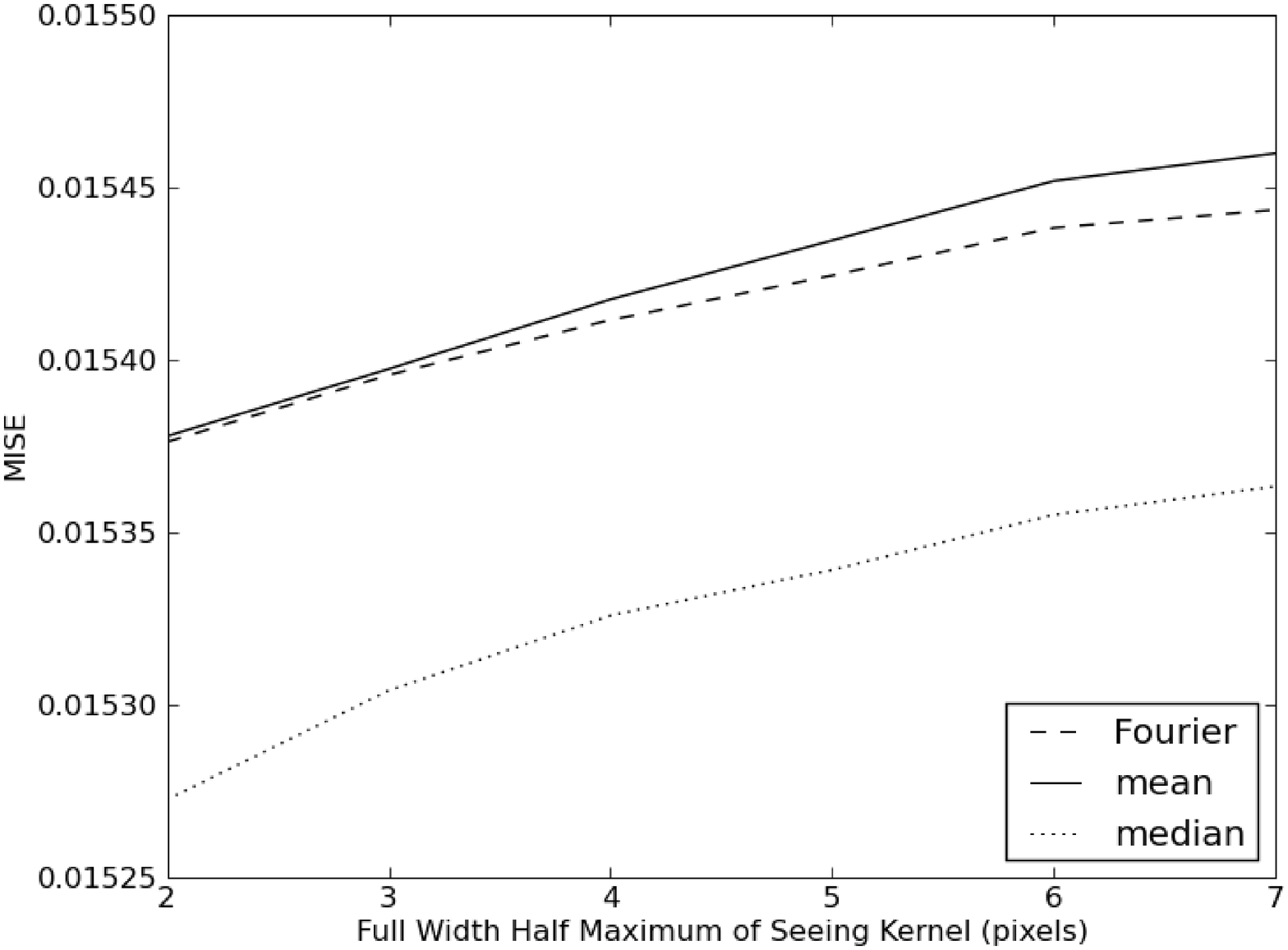}}\\
\caption{Results for the MISE comparison.  The ``Low Noise''
and ``High Noise''
  are 20:1 and 5:1 signal to noise, respectively.  Notice that the FD
method slightly outperforms the mean and median under the Gaussian
assumption.  However,
the median, being more robust to heavier-tailed distributions, dominates in the other
two regimes.}
\label{fig:results}
\end{figure}

We apply this increasing FWHM to six different noise parameterizations.
For both high variance (5:1 SNR) and low variance (20:1
SNR) we have the noise term be distributed Gaussian,
heavy-tailed shot, and inhomogeneous Poisson  noise.  In all
six combinations, a lower MISE indicates that the method conserved
the flux better, on average. See Figure \ref{fig:results} for the results. 

In all cases, the mean severity of seeing, $\mu_{FWHM}$, does not impact the
ranking of the methods.  In the Gaussian case, the methods are ranked, from best
to worst, as FD, mean, and then median.  
In both the low and high noise cases, absolute 
difference between the mean and the median is constant for all
levels of $\mu_{FWHM}$ due to the well-known efficiency gains
of the mean over the median in the Gaussian case.
However, in the high noise case, the difference in MISE
between the FD and mean methods increases 11\% as $\mu_{FWHM}$ ranges
from 2 to 7 pixels.  As the FD method uses the additional information
of a known seeing kernel, this non-constant difference is expected.

In the inhomogeneous Poisson or heavy-tailed shot noise cases, the
median outperforms both other methods.  This owes to the
particular and well known feature of the median to be more robust
against heavier-tailed distributions.  In the shot-noise case, 
the median ignores the random noise spikes with overwhelming 
probability  and we  get a noise free version of the true scene
$g$ with the median amount of seeing.  The very small advantage 
of the FD over the mean method owes to the FD's slightly 
narrower impulse response function, which we discuss in 
section \ref{sec:imageQuality} and Figure \ref{fig:expectedIRF}.
In the inhomogeneous Poisson noise case, the observed image is
created by simulating independently from a 
Poisson$\langle \theta_{ij}+\sigma^2\rangle$ from equation (\ref{eq:theta}). 
This creates a very noisy image, especially near sources, and
hence the median's ability to ignore large, transient spikes
results in the large MISE advantage.

Under all three noise distributions, the FD method displays a small but
increasing benefit in flux conservation over the mean as $\mu_{FWHM}$ goes from 2 to 7.  
It should be kept in mind that the FD method uses the known kernel assumption.   Hence,
as the seeing becomes more severe this informational advantage
should become more pronounced.  In reality, the kernel would need to be estimated and this
advantage would most likely decay.

\subsection{Image Quality}\label{sec:imageQuality}
To do this comparison we use an image where the function $g$ is a
$\delta$-function (i.e. a point source).  We use the expected
distribution of seeing at the LSST stack to draw 10 
independent and identically distributed Gaussian PSFs and
apply them to the point source.  This mimics the forward process of an
image getting blurred by the atmosphere and corresponds to operating
on $g$ with $K_i$ for $i = 1, 2, \ldots, 10$.  For this stack of
images we apply the three methods described earlier, FD,
mean, and median, to derive a
template image.  We then fit a spherical Gaussian kernel to the
template via least squares and use the diagonal element of the
covariance matrix as a measure of the width of the method's PSF.

\begin{figure}%[!h]
  \centering
  \includegraphics[width=5.0in]{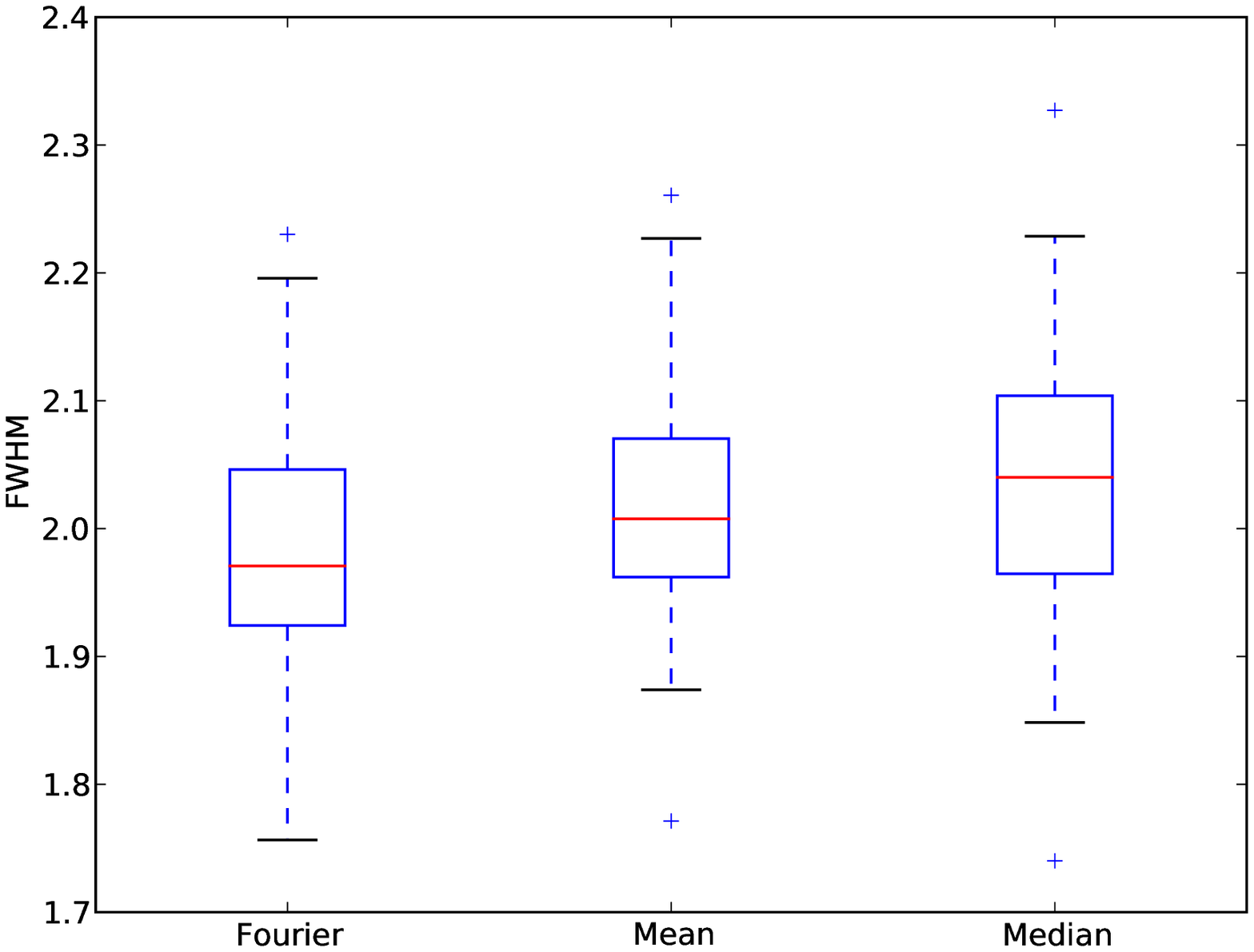}  \\
  \caption{Boxplots of the FWHM of the fitted Gaussian kernel for stacks
    created from draws from the expected seeing distribution of the LSST
    site. The three methods considered in this paper are shown.
    The boxplots are very similar and show little difference in the
    FWHM of the fitted kernels.}
\label{fig:sharpnessBoxplot}
\end{figure}

We make 1000 draws from the seeing distribution and compute this
statistic for each method and each draw.  The results are summarized in
Figure \ref{fig:sharpnessBoxplot} with boxplots of the FWHM of the
fitted Gaussian kernel.  Boxplots are a graphical tool for
quickly conveying the distribution of observed data and
are composed of a `box' and `whiskers.'
The `box' is the main rectangle, which has horizontal lines,
increasing with FWHM,
at the 25th, 50th\footnote{The 50th percentile is the median of a data set.}, 
and 75th percentiles of the data.  The `whiskers' correspond to the dotted
vertical lines with horizontal lines at the end and the plus signs.
These horizontal lines are at the 1st and 99th percentile.  The pluses are at
extreme data points.

The boxplots are very similar and show little
difference in the FWHM of the fitted kernels.  The FD method
has longer `whiskers' than the other methods, indicating a wider
range of fitted FWHM that can be expected.  It also has a very slightly
lower `box' than the other two methods, indicating that the majority
of draws result in a slightly better fitted FWHM.  For instance, 
the 50th percentile of the FD method's fitted FWHM is
1.2\% lower then the mean method's fitted FWHM.

Interestingly, one of
the draws resulted in six very good seeing images and four poor
images.  This resulted in the median method having the smallest fitted 
FWHM recorded for any method on any draw.  This is part of a more general
property of the median to behave non-continuously with the composition
of the stack.  For instance, if there were instead 4 very good seeing
images and 6 poor seeing images, the result would be a large fitted FWHM.

\subsubsection{Differences in Expectation}
The mean and FD methods are very similar in that 
they are both based around summing the images and hence
are both linear filters on the observed images.
Let's denote the estimated coadds of the two 
methods as $\hat{g}_{mean}$ and $\hat{g}_{*}$,
respectively.  We can understand the filtering that we are applying to
the true image $g$ by considering $\mathbb{E}[\hat{g}_{mean}]$ and
$\mathbb{E}[\hat{g}_{*}]$.  Note that these are
\begin{equation}
\mathbb{E}[\hat{g}_{mean}] = \left( \frac{1}{L} \sum_{i=1}^L K_i\right)g =
\int k_{mean}(x-y) g(y) dy
\qquad \textrm{and} \qquad
\mathbb{E}[\hat{g}_{*}] = \int k^{*}(x-y)g(y)dy
\label{eq:maybe}
\end{equation}
where the Fourier transform of the function $k^{*}$ is defined in the caption under
Table 1.

\begin{figure}%[!h]
  \centering
  \includegraphics[width=4.0in]{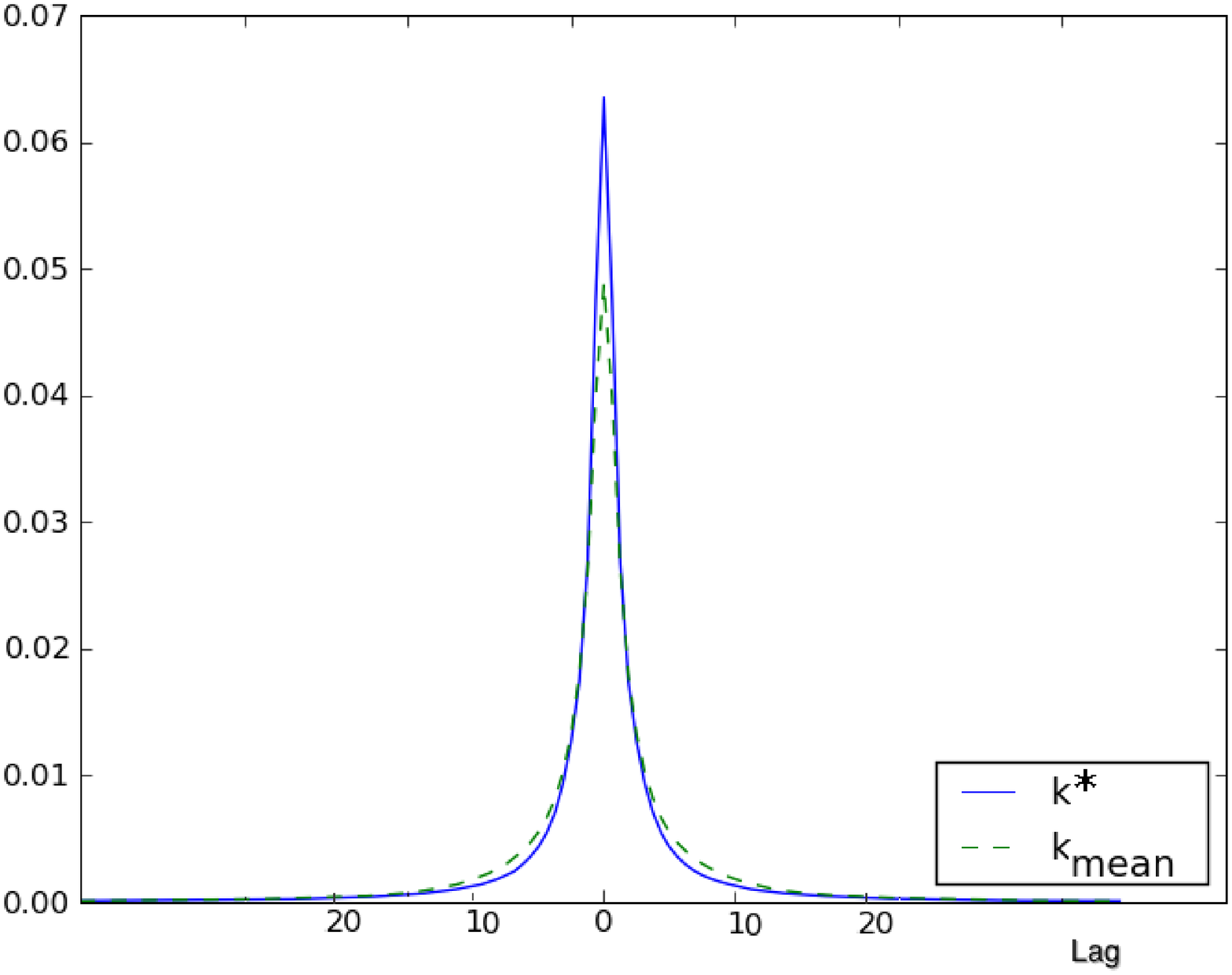}  \\
  \caption{This figure shows the IRF of the expected filter for the
    mean (dashed line) and FD (solid line)
    methods.  Though $k^*$ has more mass at lag = 0, bear in mind
    that it uses a weighting scheme that depends on knowledge of both the
    true PSFs and true variances.  Taking this into consideration, the difference is
    slight.}
\label{fig:expectedIRF}
\end{figure}

By examining the impulse
response functions (IRFs) of these filters we can find the average width
of the PSFs that will result from these methods. Note that
in this case this corresponds to the bias of each method.  
See Figure \ref{fig:expectedIRF}
for a one dimensional representation of the IRFs of the expected
filters.  The mean IRF and the Fourier
Deconvolution IRF are virtually identical and would create no visible
difference in the image quality.

%\clearpage

\section{Discussion}
A variety of template generation/image coaddition techniques have been
applied in current imaging surveys, including the CFHTLS
(\citet{gwyn2008}; mean technique), PanSTARRS
(\citet{pric2007}, mean technique), MOPEX for SIRTF images
(\citet{mako2005}, mean and median technique), and the
VIRMOS survey (\citet{rado2004}, median technique).  In this
paper, we address the broad question of whether sophisticated methods
of template generation and image coaddition are worth the added cost
and complexity, in the context of modern image surveys with large and
nearly constant data streams.  To this end, we compare Fourier domain
reconstruction techniques which have a variety of nice theoretical
properties to straightforward pixelwise statistics.  We
consider two cost metrics to evaluate the image reconstructions; Image
Quality by way of the resulting Gaussian FWHM of an image and the
conservation of flux, measured by MISE.

Under the assumptions that motivate the FD method, namely Gaussian
noise and spatially constant seeing, the FD and mean methods both have lower MISE
than the median.  However, in the heavy-tailed shot and inhomogeneous Poisson noise
cases, the median outperforms the FD method.

In situations
when there is a small amount of seeing in an image stack, corresponding
to $\mu_{FWHM} \approx 2$ pixels, we see the mean and FD methods have nearly 
the same MISE properties (less than 0.7\% difference across all 
considered noise distributions).  However, as the severity of the seeing
increases, the disparity between the methods increases as well.  As the FD
method incorporates knowledge of the seeing kernel, this property is to
be expected.  However, the  MISE for 
the high noise Gaussian, shot, and inhomogeneous Poisson cases
are only 5.4\%, 1.2\%, and 0.13\% lower, respectively, at the most 
severe levels of seeing.

These benefits are rather small, particularly in the more realistic heavy-tailed
and Poisson noise cases.
This is especially so considering the extra computational complexities
involved in the FD method over the mean and median methods.  
 
For comparing image quality of the templates, we find that the
resolution of the resulting images is very similar as the 
50th percentile of the fitted 
FWHM of the FD method is less than 2\% smaller than
the 50th percentile of the fitted FWHM of the mean method.

We find that the value added by
the FD method over the mean or median procedures does not
overcome its added cost and complexity.
This leads us to the conclusion that there is room for
improvement over all three methods.  We are currently exploring
various possible improvements and theoretical justifications for
some of the approaches outlined in this paper.

%\acknowledgments
%
%We are grateful to V. Barger, T. Han, and R. J. N. Phillips for
%doing the math in section~\ref{bozomath}.
%\email{aastex-help@aas.org <email%7Baastex-help@aas.org>}.


\end{document}